\documentclass[12pt,reqno]{article}
\usepackage{amsmath}
\usepackage{bbm}
\usepackage[hypertex]{hyperref}
\usepackage{latexsym}
\usepackage{color}

\allowdisplaybreaks \numberwithin{equation}{section}

\newcommand{\cF}{\mathcal{F}}
\newcommand{\cL}{\mathcal{L}}
\newcommand{\cV}{\mathcal{V}}
\newcommand{\tF}{D}

\newcommand{\cC}{\mathcal{C}}
\newcommand{\cG}{\mathcal{G}}
\newcommand{\cK}{\mathcal{K}}
\newcommand{\cM}{\mathcal{M}}
\newcommand{\cN}{\mathcal{N}}

\newcommand{\cT}{\mathcal{T}}
\newcommand{\cJ}{\mathcal{J}}
\newcommand{\dd}{\Delta}
\newcommand{\La}{\Lambda}
\newcommand{\Si}{\Sigma}

\newcommand{\vb}{\bar{v}}
\newcommand{\zb}{\bar{z}}
\newcommand{\rmd}{\rm d}

\newcommand{\I}{i}
\newcommand{\e}{{\rm e}}

\renewcommand{\baselinestretch}{1.2}
\newcommand{\be}{\begin{equation}}
\newcommand{\ee}{\end{equation}}
\newcommand{\bea}{\begin{eqnarray}}
\newcommand{\eea}{\end{eqnarray}}

\begin{document}
\begin{titlepage}
\begin{flushright} \small
   ITP--UU--06/05 \\ SPIN--06/03 \\ 
\end{flushright}
\bigskip

\begin{center}
\renewcommand{\thefootnote}{\fnsymbol{footnote}}
   {\LARGE\bfseries String loop corrected hypermultiplet \\[1.2ex] moduli
spaces}
\renewcommand{\thefootnote}{\arabic{footnote}}
\setcounter{footnote}{0}
\\[10mm]
\begin{center}
Daniel Robles Llana, Frank Saueressig, and
Stefan Vandoren \\[3mm]
   {\small\slshape
   Institute for Theoretical Physics \emph{and} Spinoza Institute \\
   Utrecht University, 3508 TD Utrecht, The Netherlands \\
   {\upshape\ttfamily D.RoblesLlana, F.S.Saueressig,
   S.Vandoren@phys.uu.nl} }
\end{center}
\vspace{5mm}
\end{center}

\vspace{5mm}

\hrule\bigskip

\centerline{\bfseries Abstract} \medskip \noindent
Using constraints from supersymmetry and string perturbation
theory, we determine the string loop corrections to the
hypermultiplet moduli space of type II strings compactified on a
generic Calabi-Yau threefold. The corresponding
quaternion-K\"ahler manifolds are completely encoded in terms of a
single function. The latter receives a one-loop correction
and, using superspace techniques, we argue for the existence of a
non-renormalization theorem excluding higher loop contributions.

\medskip

\hrule\bigskip
\end{titlepage}

\tableofcontents

\newpage
\section{Introduction}

The couplings of the massless modes of type II string theory
compactified on a Calabi-Yau threefold (CY$_3$) can be encoded in
low energy effective actions (LEEA) with $\cN=2$ supersymmetry.
These LEEA generally receive quantum corrections from
the world sheet conformal field theory ($\alpha^\prime$-corrections) 
and from higher genus world sheets
($g_s$-corrections). Perturbative LEEA are expanded in a double
perturbation series in $\alpha^\prime$ and $g_s$ (see e.g.\
\cite{rev1,rev2} for a review). Both from a fundamental
perspective, and in view of recent semi-realistic phenomenological
applications to $\cN=1$ theories \cite{N=1stuff}, it is important
to determine the quantum structure of such LEEA. While the
$\alpha^\prime$-corrections to the classical LEEA are well
understood, finding the perturbative $g_s$ corrections has
remained an open problem.

As the main result of this paper we determine the (one-loop) $g_s$
corrections to these LEEA. We find that the corrections are universal in
the sense that they depend on the Euler characteristic of the
CY$_3$ only. Furthermore, we argue in favor of a
non-renormalization theorem which excludes higher loop
contributions. We expect that these results also have
implications in the context of Calabi-Yau orientifold
compactifications where they could reveal new insights on the
vacuum structure.

The starting point of our investigation are $\cN = 2$, $d = 4$
supergravity actions \cite{sugrareview} coupled to vector multiplets (VM) and
hypermultiplets (HM) which provide the LEEA for type II strings
compactified on a generic CY$_3$. Supersymmetry implies that the
total moduli space $\cM$ of these theories factorizes into a local
product \cite{conformalcalculus}
\be \label{1.1} {\mathcal M} = {\mathcal M}_{\rm VM} \otimes
{\mathcal M}_{\rm HM} \, , \ee
where ${\mathcal M}_{\rm VM}$ and ${\mathcal M}_{\rm HM}$ are
parameterized by the scalars of the VM and HM, respectively.
Supersymmetry further dictates that ${\mathcal M}_{\rm VM}$ be a
special K\"ahler (SK) manifold \cite{deWit:1984pk}, and ${\mathcal
M}_{\rm HM}$ a quaternion-K\"ahler (QK) manifold \cite{BW}. 
For compactifications
of the type IIA string the volume modulus sits in a VM while the
dilaton is in the HM sector.\footnote{By this, we mean the four-dimensional
dilaton in which the factorization property \eqref{1.1} holds. 
We will explain its relation to the string coupling constant in later
sections.} This implies that ${\mathcal M}_{\rm
VM}$ gets $\alpha^\prime$-corrections while ${\mathcal M}_{\rm
HM}$ receives $g_s$-corrections only. Compactifying type IIB
strings the volume modulus and the dilaton are both in the HM
sector. Hence the HM sector receives both $\alpha^\prime$ and
$g_s$-corrections while the  VM sector is classically exact.

The type IIB VM prepotential can be computed through knowledge of
the Yukawa couplings (or triple intersection forms) for the given
CY$_3$. Applying mirror symmetry this result can then be used to
determine the VM couplings in the type IIA compactification
including $\alpha^\prime$-corrections \cite{mirrorsymmetry} (for a
review, see \cite{msrev}). This gives, at least in principle, the
complete picture of the VM moduli space in these
compactifications.

The corresponding picture in the HM sector is, however, less
complete. This is mainly due to the lack of a (non-)perturbative
duality that relates a classically exact sector of the M-theory
moduli space to $\cM_{\rm HM}$.\footnote{In principle, one could
use the string-string duality between heterotic strings
compactified on $T^2 \times K3$ and type IIA string theory on a
$K3$-fibered CY$_3$ to obtain the fully quantum corrected result.
Even though there are no $g_s$ corrections to the HM moduli space
on the heterotic side, this space remains poorly understood
already at the classical level.} The classical result for
$\cM_{\rm HM}$ can be obtained by the (classical) c-map
\cite{CFG,FS} which relates the VM sector of the type IIA (IIB) to
the HM sector of the type IIB (IIA) string compactification on the
same CY$_3$. But beyond this classical result only little is known
about string loop corrections to this sector. While we will
elaborate on the perturbative corrections below, non-perturbative
corrections due to D-brane and NS5-brane instantons have been
proposed in \cite{BBS}. (See \cite{Davidse:2004gg,Membrane} for
some results about such instanton corrections to the universal
hypermultiplet.)

In this paper we will determine the form of the one-loop
corrections to the HM moduli space in a generic CY$_3$
compactification of type II strings. Instead of doing this by
performing explicit string loop calculations, we will impose the
constraints from ${\cal N}=2$ supersymmetry, together with generic
properties that string perturbation theory has to satisfy. 
Our starting point is the
list of Strominger's \cite{Strominger} which summarizes the
properties of the perturbatively corrected HM moduli spaces:
\begin{enumerate}
\item due to $\cN = 2$ supersymmetry the quantum-corrected metrics should be quaternion-K\"ahler,
\item the corrections to the classical result should be subleading in the dilaton ($g_s$),
\item the Peccei-Quinn symmetries (cfg.\ eq.\ \eqref{eq:PQ})
are preserved at the perturbative quantum level,
\item since string amplitudes with an odd number of RR fields vanish, the perturbations to the classical result always contain an even number of RR fields,
\item parity is a symmetry,
\item and the full perturbatively corrected metrics should be consistent with the known results from string loop computations \cite{HM1,AntoniadisUH}.
\end{enumerate}
These conditions turn out to be sufficient to determine the HM metric, 
and our main result is given in
eq. \eqref{finallagrangian}, together with \eqref{identff} in type
IIA, and \eqref{identffb} in type IIB.

In \cite{AntoniadisUH} these conditions have been implemented for
the case of the universal hypermultiplet with the result that this
sector receives non-trivial quantum corrections proportional to
the Euler characteristic of the (rigid) CY$_3$. Subsequently, this
result has been rewritten in superspace, in terms of a single
function that determines all the components of the one-loop
corrected moduli space metric \cite{UHMsuper}. The implementation of
Strominger's list on QK metrics of arbitrary dimension has been
considered in \cite{Gunther,Gunther2}, but remained inconclusive
due to technical problems in enforcing the QK condition on the
deformations of the classical HM moduli space.

In this paper we use the off-shell formulation of superconformal
tensor multiplets \cite{Tensor1,HKC1} to determine the
perturbative corrections to $\cM_{\rm HM}$. The main advantage of
the off-shell formulation is that one can describe QK metrics, and
therefore the effective action, in terms of a single function.
Further simplifications occur when there are additional
isometries, like the Peccei-Quinn symmetries present in string
perturbation theory. For $4n$-dimensional QK metrics with $n+1$
commuting shift symmetries one can use the duality between hyper-
and tensor multiplets (TM) in four dimensions. In that case the
off-shell description can be given in terms of ${\cal N}=2$ tensor
multiplets and is also known \cite{Rocek1} to be
determined by a single function which we will call $H$ in the following.

At the classical level $H$ has recently been constructed in
\cite{RVV}, see also \cite{Berkovits:1995cb,Berkovits:1998jh} for
related results. This function was found by describing the c-map
\cite{CFG,FS} off-shell. We here search for deformations of this
map which satisfy the conditions (1) - (6) stated above, and find
the general solution. The deformed functions $H$ thus provide a
quantum c-map which determine the perturbative corrections to the
QK metrics arising from generic CY$_3$ compactifications of the
type II string.

The rest of the paper is organized as follows. In Section 2 we
review the facts about supergravity theories and the classical
c-map which are relevant in our construction. Section 3 outlines
the superconformal quotient for the superconformal TM lagrangian
which is then applied to the classical c-map. In Section 4 the
one-loop corrections and the resulting HM moduli spaces are
constructed and in Section 5 we argue for a non-renormalization
theorem excluding higher loop corrections. Section 6 contains some
discussion and an outlook. The technical details of our
constructions are collected in Appendix A.
%
\section{Effective supergravity actions}

In this section we describe the four-dimensional $\cN = 2$
supergravity actions that provide the LEEA for type II strings
compactified on a generic CY$_3$. We start be reviewing their tree
level moduli spaces together with the c-map in Subsection
\ref{sugra}. Subsection \ref{off-shell}
discusses the off-shell formulation of Poincar\'e supergravity based on the superconformal calculus which will play a central role in our construction.

\subsection{Tree level effective actions}
\label{sugra}

Type II string compactifications on CY$_3$ yield four-dimensional
$\cN = 2$ supergravity theories including $n_V$ vector
and $n_H$ hypermultiplets (or, equivalently, their  tensor
multiplet duals, see below). Denoting the Hodge numbers of the
CY$_3$ by $h^{1,1}$ and $h^{2,1}$, compactifications of type IIA strings
yield $n_V =h^{1,1}$, $n_H=h^{2,1}+1$. In the type IIB case the
Hodge numbers are interchanged, i.e., $n_V=h^{2,1}$,
$n_H=h^{1,1}+1$. $\cN = 2$ supersymmetry further requires that the
scalar manifolds of the theory factorize according to \eqref{1.1}.

The VM scalars parameterize the manifold $\cM_{\rm VM}$ which is
local (projective) special K\"ahler
   \cite{deWit:1984pk,conformalcalculus}. These manifolds are characterized by a so-called prepotential, a holomorphic function $F(X^\Lambda)$, $\Lambda = 1, \ldots, n_V + 1$, which is homogeneous of degree two in $X^\Lambda$. Introducing projective coordinates
\be \label{2..1}z^\Lambda = \frac{X^\Lambda}{X^1} = \left\{ 1 ,
z^a \right\} \, , \quad a = 1, \ldots , n_V \, , \ee
the metric and K\"ahler
potential of $\cM_{\rm VM}$ are given by
\be\label{eq:KM} \cG_{a {\bar b}} =
\partial_{a} \partial_{{\bar b}} \cK \ ,\qquad
{\mathcal K} (z,\bar z) = \ln \, (z^\Lambda N_{\Lambda\Sigma} \bar
z^\Sigma) \ , \ee
in which $N_{\Lambda\Sigma} =\I \left(F_{\Lambda\Sigma} - \bar{F}_{\Lambda\Sigma} \right)$ and
$F_\Lambda(X) = \frac{\partial}{\partial X^\Lambda} F(X)$ etc.
Furthermore, the kinetic terms of the VM gauge fields are
determined by the matrix
\be\label{eq:cN} {\mathcal N}_{\Lambda\Sigma} = -i\bar
F_{\Lambda\Sigma} - \frac{(N z)_\Lambda (N z )_\Sigma}{(z N z)}\ ,
\ee
where $(Nz)_\Lambda = N_{\Lambda\Sigma} z^\Sigma$ and $ (zN{z})=
z^\Lambda N_{\Lambda\Sigma} { z}^\Sigma$. When considering CY$_3$
compactifications the classical part of $F(X^\Lambda)$ is
determined by the triple-intersection numbers of the CY$_3$. In
compactifications of the type IIA string the prepotential
additionally receives perturbative and non-perturbative
$\alpha^\prime$ corrections.

The HM scalars parameterize the manifold ${\mathcal M}_{\rm HM}$
which must be quaternion-K\"ahler \cite{BW}. At tree level in the
string coupling constant, the
corresponding HM lagrangians for the type IIA (IIB)
compactification are related to the special K\"ahler geometry of
the IIB (IIA) compactification on the same CY$_3$ via the c-map
\cite{CFG,FS}.\footnote{Without gravity, one can also perform a (rigid) c-map, that maps vector multiplets to hypermultiplets. In terms of geometries, the
map is between rigid special K\"ahler spaces and hyperk\"ahler spaces 
\cite{CFG,DeJaegher:1997ka,Para}.} Alternatively, refs.\ \cite{BCF,BGHL} derived these
lagrangians from a classical compactification of ten-dimensional
IIA or IIB supergravity on a generic CY$_3$. The bosonic part of
the resulting hypermultiplet lagrangians can then be written as (in conventions
with $\kappa^{-2}=2$)
\be \label{eq:cmap}
\begin{split}
e^{-1} \cL = &\,- R - \tfrac{1}{2}( \partial_\mu \phi )^2
+2\cG_{a{\bar b}}
\partial_\mu z^a \partial^\mu {\bar{z}}^{\bar b}
   +\tfrac{1}{2} {\rm e}^{-\phi} \left( \cN + \cN \right)^{\Lambda\Sigma}
| 2 \cN_{\Lambda \Pi} \partial_\mu A^\Pi + i \partial_\mu B_\Lambda |^2 \\
& \, -\tfrac{1}{2} {\rm e}^{-2 \phi} \left( \partial_\mu \sigma -
\tfrac{1}{2} \left( A^\Lambda \partial_\mu B_\Lambda - B_\Lambda
\partial_\mu A^\Lambda \right) \right)^2 \, .
\end{split}
\ee
Here $\phi$ is the four-dimensional dilaton, $\sigma$ is the dual
scalar arising from the NS two-form, $z^a$, $a=1,...,n_H-1$, are
the geometric moduli (i.e., complex structure or K\"ahler moduli in IIA or IIB,
respectively) and the $2n_H$ additional real scalars $A^\Lambda,
B_\Lambda$ arise from compactifying the (ten-dimensional) RR
fields. Furthermore, $\cG_{a \bar{b}}$ and ${\mathcal
N}_{\Lambda\Sigma}$ are the metric \eqref{eq:KM} and gauge kinetic
matrix \eqref{eq:cN} of the dual special K\"ahler geometry, and
$({\mathcal N}+\bar{\mathcal N})^{\La\Si}$ is the inverse of
$({\mathcal N}+\bar{\mathcal N})_{\La\Si}$. Note that
\eqref{eq:cmap} is completely fixed by the underlying prepotential
$F(X^\Lambda)$. In particular, this allows to determine the string
tree-level $\alpha^\prime$-corrections to the type IIB
hypermultiplet geometry by substituting the
$\alpha^\prime$-corrected VM prepotential of the IIA
compactification.

The compactification of the ten-dimensional tensor fields
naturally induces certain symmetries in the resulting LEEA which have been studied
 in detail in ref.\ \cite{deWit}. For our purpose it suffices to note that
the lagrangian \eqref{eq:cmap} is invariant under the $2 n_H + 1$
Peccei-Quinn symmetries
\be\label{eq:PQ}
\begin{split}
\delta \sigma = \epsilon + \tfrac{1}{2} \left( \alpha^\Lambda
B_\Lambda - \beta_\Lambda \, A^\Lambda \right) \; , \quad \delta
A^\Lambda = \alpha^\Lambda \; , \quad \delta B_\Lambda =
\beta_\Lambda \, ,
\end{split}
\ee
where $\epsilon, \alpha^\Lambda$ and $\beta_\Lambda$ are $2 n_H +
1$ real parameters. These isometries constitute a $2 n_H +
1$-dimensional Heisenberg algebra. Since these isometries
originate from tensor fields in ten dimensions, this algebra is
expected to be preserved at the perturbative quantum level
\cite{Strominger}.

These shift symmetries suggest that there should be a description
of the lagrangian \eqref{eq:cmap} in terms of tensor multiplets.
In fact such a description naturally arises in compactifications
of type II strings. For type IIA, one obtains $h^{2,1}$
hypermultiplets and one tensor multiplet. The latter can be
dualized into a scalar yielding hypermultiplets only.
Compactifying type IIB strings yields a double-tensor multiplet
\cite{TV1} and $h^{1,1}$ tensor multiplets. The bosonic part of
the lagrangian for this system was found in \cite{BGHL}, and reads
\be \label{2.6}
\begin{split}
e^{-1}{\mathcal L}^{\rm TM}_{\rm cl} = & \, - \frac{1}{2}
(\partial_\mu\phi)^2 + 2 \, {\mathcal G}_{a \bar b}\partial_\mu
z^a \partial^\mu \zb^{\bar b} + \frac{1}{2}{\rm e}^{-\phi}
({\mathcal N}+\bar{\mathcal N})_{\La\Si}
\partial_\mu A^\La\partial^\mu A^\Si \\
& \, + 2 \, \cT^{\rm cl}_{IJ} \, E_\mu^I E^{J\mu} + i({\mathcal
N}-\bar{\mathcal N})_{\La\Si}\left[(\partial_\mu
A^\La)E^{\Si\mu}-2(\partial_\mu A^\La)A^\Si E^{0\mu}\right] \, .
\end{split}
\ee
Here \footnote{We use Pauli-K\"all\'en conventions where
$\varepsilon^{0123} = i$, so $E^\mu$ is real.} $E^{\mu} =
\tfrac{i}{2} e^{-1} \varepsilon^{\mu \nu \rho \sigma} E_{\rho
\sigma}$ is the field strength of the antisymmetric tensor field
$E_{\mu \nu}$. The index $I$ runs over one more value than
$\Lambda$, so $I=\{0,\Lambda\}$. This is because, compared to
\eqref{eq:cmap}, both $B_I$ and $\sigma$ have been exchanged for
tensors. The matrix $\cT^{\rm cl}_{IJ}$ appearing in the tensor
field kinetic term is given by
\be \label{2.7} \cT^{\rm cl}_{IJ}  = {\rm e}^{\phi} \left[
\begin{array}{cc} {\rm e}^{\phi} - ({\mathcal N}+\bar {\mathcal
N})_{\La\Si}A^\La
A^\Si & \frac{1}{2}({\mathcal N}+\bar{\mathcal N})_{\La\Si}A^\La \\
\frac{1}{2}({\mathcal N}+ \bar{\mathcal N})_{\La\Si}A^\Si &
-\frac{1}{4}({\mathcal N}+\bar{\mathcal
N})_{\La\Si}\end{array}\right] \, . \ee
Dualizing the tensor field strengths back into scalars, one
obtains the hypermultiplet lagrangian \eqref{eq:cmap}. This will
be done explicitly in Subsection \ref{classexample}.

\subsection{Off-shell formulation}
\label{off-shell}
As will become clear in this paper, working in an off-shell
formalism turns out to be advantageous. The main reason is that
the matter sectors in \eqref{eq:cmap} and \eqref{2.6} can be
elegantly  recast into superspace language. Having such a
formulation at hand considerably simplifies addressing the
question of the loop corrections in the following.

Instead of considering the standard Poincar\'e supergravity
described above, this off-shell formulation utilizes a gauge
equivalent formulation based on the
superconformal group. The multiplet containing
   the gravitational degrees of freedom in this locally
superconformal invariant theory is called the Weyl multiplet. It
contains the graviton and gravitinos as well as gauge fields for
the $U(1)$ and $SU(2)$ $R$-symmetry groups that belong to the
bosonic part of the superconformal group. Moreover, the theory can
include any number of vector, hyper-, and tensor multiplets whose
superconformal couplings have been worked out in
\cite{scactions1,conformalcalculus,superconformalhypers} and
\cite{Tensor1}. In order to gauge fix to Poincar\'e supergravity
one needs at least one vector and one hypermultiplet, which can
act as compensators for the extra symmetries of the theory.
Alternatively, as we will use later, the hypermultiplet
compensators can be replaced by four compensating scalars in two
tensor multiplets. Eliminating the auxiliary $U(1)$ and $SU(2)$
gauge fields, combined with appropriate gauge fixing conditions
yields the Poincar\'e theory in which the moduli spaces ${\mathcal
M}_{\rm VM}$ and ${\mathcal M}_{\rm HM}$ or ${\mathcal M}_{\rm
TM}$ appear. This is the basic idea of the ${\cal N}=2$
superconformal calculus \cite{scc1} and the gauge fixing procedure
is called the superconformal quotient.

\begin{figure}[t]
\begin{center}
\setlength{\unitlength}{1cm}
\begin{picture}(13.5,4)
\thicklines \put(0,3){\makebox(1,1){$\cM^{\rm SC}_{\rm TM}$}}
\put(6,3){\makebox(1,1){$\cM^{\rm SC}_{\rm HM}$}}
\put(8,3){\makebox(1,1){$\otimes$}}
\put(10,3){\makebox(1,1){$\cM^{\rm SC}_{\rm VM}$}}
\put(1.5,3.5){\vector(1,0){4}} \put(5.5,3.5){\vector(-1,0){4}}
\put(0,0){\makebox(1,1){$\cM^{\rm  }_{\rm TM}$}}
\put(6,0){\makebox(1,1){$\cM^{\rm  }_{\rm HM}$}}
\put(8,0){\makebox(1,1){$\otimes$}}
\put(10,0){\makebox(1,1){$\cM^{\rm  }_{\rm VM}$}}
\put(1.5,0.5){\vector(1,0){4}} \put(5.5,0.5){\vector(-1,0){4}}
\put(0.5,1){\vector(0,1){2}} \put(0.5,3){\vector(0,-1){2}}
\put(6.5,1){\vector(0,1){2}} \put(6.5,3){\vector(0,-1){2}}
\put(10.5,1){\vector(0,1){2}} \put(10.5,3){\vector(0,-1){2}}
\put(11.2,1.7){\makebox(3,1){superconformal}}
\put(11.2,1.2){\makebox(3,1){quotient}}
\put(2,-0.7){\makebox(3,1){scalar-tensor duality}}
\put(2,3.7){\makebox(3,1){scalar-tensor duality}}
\end{picture}
\end{center}
\renewcommand{\baselinestretch}{1}
\caption{\label{p.1}{\footnotesize Relations between the scalar
manifolds featuring in the conformal (top) and Poincar\'e
supergravity (bottom). The vertical arrows indicate that these
theories are related by the superconformal quotient while the
horizontal arrows imply that, provided the HM scalar manifolds
have suitable isometries, scalars are dual to antisymmetric tensor
fields thus relating the corresponding hyper- and tensor multiplet
scalar manifolds.}}
\end{figure}
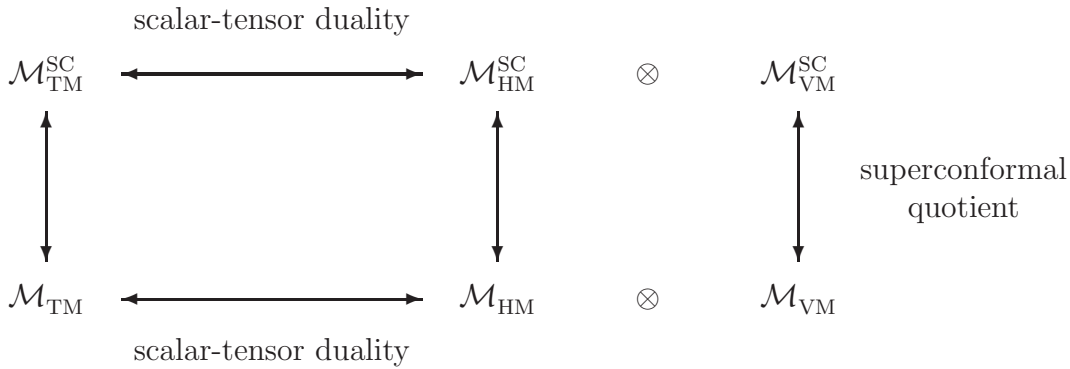

The scalars of the vector, hyper- and tensor multiplets featuring
in the superconformal theory parameterize the scalar manifolds
${\mathcal M}_{\rm VM}^{\rm SC}$, ${\mathcal M}_{\rm HM}^{\rm SC}$
and ${\mathcal M}_{\rm TM}^{\rm SC}$, respectively. The
superscript ``SC'' indicates that the corresponding
manifolds characterize a superconformal theory. The relations
between these manifolds and their counterparts in Poincar\'e
supergravity  are summarized in Fig.\ \ref{p.1}. It turns out that
$\cM^{\rm SC}_{\rm VM}$ is a rigid (affine) special K\"ahler
manifold of real dimension $2n_V + 2$. Its metric is completely
determined by the holomorphic prepotential $F(X^{\Lambda})$
homogeneous of second degree. This prepotential defines the
K\"ahler potential $K$ and the metric $N_{\Lambda \Sigma}$ on
$\cM_{\rm VM}^{\rm SC}$ as
\be \label{2.3} K = i \left( \bar{X}^{\Lambda} F_\Lambda(X) -
X^{\Lambda} \bar{F}_\Lambda(\bar{X}) \right) \, , \qquad
N_{\Lambda \Sigma} = i \left( F_{\Lambda \Sigma} -
\bar{F}_{\Lambda \Sigma} \right) \, . \ee
Taking the superconformal quotient of $\cM_{\rm VM}^{\rm SC}$ then
leads to the local special K\"ahler manifold $\cM_{\rm VM}$ of real
dimension $2n_V$ detailed above. Furthermore, while ${\mathcal
M}_{\rm HM}$ is quaternion-K\"ahler, it turns out that its
superconformal counterpart ${\mathcal M}_{\rm HM}^{\rm SC}$ is a
hyper-K\"ahler cone whose geometrical properties are completely
specified by a single function, the hyper-K\"ahler potential
\cite{Swann,deWit:1998zg}. The relation between hyper-K\"ahler cones
and quaternion-K\"ahler manifolds was studied in more detail in
\cite{HKC1,HKC2,HKC3,Bergshoeff:2004nf}. Furthermore, this map was utilized
to construct LEEA for CY$_3$ compactifications undergoing flop and
conifold transitions in \cite{flop}.

For the purpose of this paper, the most convenient starting point
is the superconformal TM lagrangian \cite{Tensor1}. Building on
earlier work \cite{Rocek1}, it turns out that the corresponding
scalar geometry $\cM^{\rm SC}_{\rm TM}$ is encoded by a single function
$\cF(x,v,\vb)$, which completely
determines the lagrangian. This function can be expressed in terms of a
contour integral,
\be\label{eq:contour} {\mathcal F}(v^I,\vb^I,x^I)=-{\rm
Im}\left[\frac{1}{2\pi i}\oint_{\mathcal C}\frac{\rmd
\zeta}{\zeta} H(\eta^{I})\right]\ . \ee
Here $I,J = 0, \ldots , n_H+1$ enumerates the tensor multiplets and
the three scalars of each tensor multiplet appear in the combination
\be\label{etadef} \eta^I
= \frac{v^I}{\zeta} +  x^I - \zeta \vb^I \, .
\ee
The contour integral representation guarantees that the function
${\cal F}$ satisfies the following differential equation
\cite{Gates:1984nk,Rocek1}
\be\label{eq:Fconst}
\cF_{x^I x^J} + \cF_{v^I \vb^J} = 0 \, .
\ee
Conformal invariance requires the function $H$ to be
homogeneous of degree one~\footnote{This homogeneity
has to be understood under the contour integral. Linear terms of
the form $H\propto \eta$ vanish in the final lagrangian, while
terms of the form $H\propto \eta {\rm ln} \eta$ are non-vanishing,
but only homogenous of degree one up to terms that vanish in the
final lagrangian. For more details, see e.g. \cite{HKC1}.} under
rescalings of $\eta^I$ and have no explicit $\zeta$ dependence
while $\cC$ is a curve in the complex $\zeta$ plane. All this
naturally follows from $\cN=2$ projective superspace
\cite{Gates:1984nk,Karlhede:1984vr}, in which the the $v^I$ are
the scalars coming from a ${\mathcal N}=1$ chiral superfield while
the $x^I$ are the real scalar fields of an ${\mathcal N}=1$ tensor
multiplet. Together they compose to an $\cN = 2$ tensor multiplet.
The tensor multiplet sector of the conformal supergravity theory
is then completely specified by the function $H(\eta)$ which,
besides being homogeneous of degree one, does not need to satisfy
any further constraints. The rigidly superconformal tensor multiplet
lagrangian is described by integrating the function ${\cal F}$ over
the ${\cal N}=2$ projective superspace measure. The coupling to
the Weyl multiplet is described in the next section.

The question then arises what is the function $H$ that, upon
taking the superconformal quotient, gives rise to the lagrangians
\eqref{eq:cmap} or \eqref{2.6}. This was recently answered
in \cite{RVV}, where it was shown that
\be \label{eq:confc} H^{\rm cl}(\eta) =
\frac{F(\eta^\Lambda)}{\eta^0} \, . \ee
Here $F(X)$ is the holomorphic prepotential determining the VM
sector of the dual type II compactification, but now evaluated as
a function of the TM fields $\eta^\Lambda$ while $\eta^0$ is an
additional compensator. To establish that \eqref{eq:confc} indeed
gives rise to the hypermultiplet lagrangian \eqref{eq:cmap}, it is
convenient to work in a partially gauge-fixed version where $v^0 =
\vb^0 = 0$. Then the contour $\cC$ can be taken around the pole
$\zeta = 0$ in the complex $\zeta$-plane. Assuming that $F(\zeta
\eta^\Lambda)$ has no poles at $\zeta = 0$, the contour integral
yielding the function $\cF(v,\vb,x)$ can then easily be evaluated.
By dualizing the resulting conformal tensor multiplet theory to
hypermultiplets and subsequently performing the superconformal
quotient on the resulting hyper-K\"ahler cone $\cM^{\rm SC}_{\rm
HM}$ it was verified that the resulting metrics on $\cM_{\rm HM}$
are indeed given by \eqref{eq:cmap} \cite{RVV}. We rederive this
result in the next section.

Complementary to the classical result \eqref{eq:confc}, ref.\
\cite{UHMsuper} constructed the function $H(\eta)$ encoding the
one-loop corrected universal hypermultiplet lagrangian found by
Antoniadis et.\ al.\ \cite{AntoniadisUH}. With $F(\eta) = -i
(\eta^1)^2$ describing the classical  universal hypermultiplet,
the one-loop corrected metric can be obtained from
\be \label{qcuhm} H^{\rm UHM}(\eta) = -i \frac{(\eta^1)^2}{\eta^0}
+ 4 \, i \, c \, \eta^0 \, \ln(\eta^0) \, , \ee
where $c$ is an a priori undetermined constant. In the gauge $v^0 = \vb^0 = 0$ the contour $\cC$
is taken around the origin. Alternatively the contour integral can be evaluated without making this partial gauge choice by choosing a different contour \cite{UHMsuper}. 

That the second term describes a one-loop term of order $g_s^2$ higher than
the classical term can be understood as follows. The string coupling 
is a dimensionless quantity. The tensor multiplets have scaling dimensions, 
so only a ratio can be proportional to $g_s$. From the explicit
calculation in \cite{UHMsuper}, it follows that 
$\eta^1/\eta^0$ scales like $g_s^{-1}$. It is then easy to see that the second
term is of order $g_s^2$ higher.

It should be clear then that the problem of how to incorporate
string loop corrections to the lagrangians \eqref{eq:cmap} and
\eqref{2.6} is most easily done in the off-shell description.
Combined with \eqref{eq:confc}, it will turn out to be easy to
generalize \eqref{qcuhm} to the case of more hypermultiplets. We
discuss this in Section 4.

\section{Tensor multiplet lagrangians}
\label{tensorquotient}

Before we discuss the loop corrections, we first construct the 
most general ${\cal N}=2$ supergravity action coupled to tensor multiplets in 
components \cite{Tensor1}. This will enable us to compare with known 
results for string loop amplitudes to the effective action, which we discuss
in the next section.

The natural starting point for our investigation is the
superconformal tensor multiplet lagrangian \cite{Tensor1}
including $n_H + 1$ tensor multiplets. This was also the starting
point for constructing the classical conformal c-map \cite{RVV}
where upon determining the function $\cF(v,\vb,x)$, the
hyper-K\"ahler potential of the corresponding hypermultiplet
geometry $\cM^{\rm SC}_{\rm HM}$ was found as the Legendre
transform of $\cF(v,\vb,x)$ with respect to $x^I$. Subsequently
the superconformal quotient was taken on the hypermultiplet side
along the lines of \cite{HKC1}.

Including a logarithmic correction of the form \eqref{qcuhm},
however, this strategy faces the obstacle that the Legendre
transform of $\cF(v,\vb,x)$ with respect to $x^0$ involves solving
a transcendental equation which cannot be done explicitly. To
avoid this complication we take a different route through Fig.\
\ref{p.1} and first take the superconformal quotient on the tensor
side before dualizing the tensors to scalar fields. The
superconformal quotient for TM is then subject of Subsection
\ref{TMscq}. In Subsection \ref{classexample} we utilize this
formalism to derive the classical hypermultiplet lagrangian
\eqref{eq:cmap} starting from eq.\ \eqref{eq:confc}.

\subsection{The superconformal quotient for tensor multiplets}
\label{TMscq}

We start by considering $n_H + 1$ tensor multiplets which are
conformally coupled to the Weyl multiplet. The bosonic degrees of
freedom of the $\cN = 2$ tensor multiplet \cite{ITM} consist of an antisymmetric tensor
field $E_{\mu \nu}$ with field strength $E^\mu := \frac{\I}{2}
e^{-1} \varepsilon^{\mu \nu \rho \sigma} \partial_\nu E_{\rho
\sigma}$, an $SU(2)$ triplet of scalars $L^{ji } = L^{ij} =
(L_{ij})^*$, $i,j = 1,2$, satisfying the reality condition
$L^{kl}= \varepsilon^{ki} \varepsilon^{l j}\,L_{ij}$, and a
complex auxiliary field $G$ which will play no role in the
following. The bosonic part of the Weyl multiplet contains the
vielbein $e_\mu{}^a$, an auxiliary field $D$, and the
(non-dynamical) gauge fields $\vec{\cV}_{\mu}$, $A_\mu$, $b_\mu$
gauging the $SU(2)$, $U(1)$ and dilatations of the superconformal
algebra. Furthermore, we have a dependent gauge field $
f_{\mu}{}^{\mu} = \tfrac{1}{6} R - D $ which is related to special
conformal transformations.

In order to make contact with the tensor multiplet geometry
outlined in Subsection \ref{sugra} we decompose the $L^{ij I}$ as
\be  \label{eq:L-xv}
    L^{12\,I} \equiv \tfrac1{2} \I \,x^I\,, \qquad
    L^{11\,I}\equiv v^I\, , \qquad L^{22\,I}\equiv \vb^I\, .
\ee
In these coordinates the bosonic part of the superconformal tensor
multiplet lagrangian \cite{Tensor1} can be concisely written as
\be\label{1.3}
\begin{split}
e^{-1} \cL = & \, \cF_{x^I x^J} \left( \partial_\mu v^I
\partial^\mu \vb^J + \tfrac{1}{4} \partial_\mu x^I \partial^\mu
x^J - E^I_\mu \, E^{J \mu}\right)
+ \I \, E^I_\mu \left( \cF_{v^I x^J} \, \partial^\mu v^J - \cF_{\vb^I x^J} \partial^\mu \vb^J \right) \; \;  \\
& \, +  \tfrac{1}{2} \left( \vec{\cV}_{\mu} \right)^{\rm T} \, M
\, \vec{\cV}^{\mu} + \vec{\cV}_{\mu} \cdot \left( \vec{S}^{\mu} +
\vec{T}^{\mu} \right)
   - 2 \cF_{x^I x^J} \left( v^I \vb^J + \tfrac{1}{4} x^I x^J
\right) \left( \tfrac{1}{3} R + D \right) \, .
\end{split}
\ee
The $SU(2)$ triplet of gauge fields $\vec{\cV}_{\mu}$ couples to
the $SU(2)$ currents
\be \label{vecS} \vec{S}_{\mu} = - \, \frac{\I}{2} \cF_{x^I x^J}
\left[
\begin{array}{c}
v^I \, \partial_{\mu} x^J - x^I \, \partial_\mu  v^J \\
- \, \vb^I \, \partial_{\mu} x^J + x^I \, \partial_\mu  \vb^J \\
2 \left( \vb^I \, \partial_{\mu} v^J - v^I \, \partial_\mu  \vb^J
\right)
\end{array}
\right] \, , \ee
and
\be \label{vecT} \vec{T}_{\mu} = \cF_{x^I x^J}  E^I_\mu \left[ v^J
\, , \, \vb^J \, , \, x^J \,  \right]^{\rm T} \, , \ee
containing derivatives of the scalar and tensor fields,
respectively. Furthermore, the matrix $M$ which appears in the
term quadratic in $\vec{\cV}_\mu$ is, in the canonical complex
basis $\vec{\cV}_\mu=({\cV}_\mu^+,{\cV}_\mu^-,{\cV}_\mu^3)$ given
by
\be\label{1.2} M = \left[
\begin{array}{ccc}
- \tfrac{1}{2} \cF_{x^I x^J} v^I v^J & \tfrac{1}{2} \cF_{x^I x^J}
\left( v^I \vb^J + \tfrac{1}{2} x^I x^J \right) &
- \tfrac{1}{2} \cF_{x^I x^J} x^I v^J \\
\tfrac{1}{2} \cF_{x^I x^J} \left( v^I \vb^J + \tfrac{1}{2} x^I x^J
\right) & - \tfrac{1}{2} \cF_{x^I x^J} \vb^I \vb^J &
- \tfrac{1}{2} \cF_{x^I x^J} x^I \vb^J \\
- \tfrac{1}{2} \cF_{x^I x^J} x^I v^J & - \tfrac{1}{2} \cF_{x^I
x^J} x^I \vb^J &
2 \cF_{x^I x^J} \vb^I v^J\\
\end{array}
\right] \, . \ee
Observe that the lagrangian \eqref{1.3} and in particular the
metric on $\cM^{\rm SC}_{\rm TM}$ is completely fixed by
specifying the function $\cF(v, \vb, x)$ which is subject to the
conditions \eqref{eq:Fconst}.

The superconformal quotient is performed by making a gauge choice
for the $SU(2)$ symmetry and dilatations together with eliminating
the gauge fields $\vec{\cV}_{\mu}$ and the auxiliary field $D$ by
their equations of motion. For the fields $\vec{\cV}_{\mu}$ this
is straightforward. Here we can use the freedom of performing
$SU(2)$ rotations to fix
\be\label{eq:SU2-eq} v^0 = 0 \; , \quad v^1 = \vb^1 \, , \ee
and then eliminate the non-dynamical fields $\vec{\cV}_{\mu}$.

The consistent elimination of $D$ is slightly more complicated and
requires introducing a conformal vector multiplet which provides
the compensator for the $U(1)$ symmetry acting in the vector
multiplet sector. The relevant piece coming from the conformal
vector multiplet lagrangian can readily be taken from
\cite{Tensor1} and reads\footnote{Since we are interested in
quantum corrections to the hypermultiplet sector, we will ignore
the vector multiplet geometry in the following and  include the
terms required for gauge fixing the superconformal tensor
multiplet theory only. It is, however, straightforward to also
include the complete vector multiplet sector in the setup.}
\be\label{eq:VM} e^{-1} \cL^{\rm VM} = - i \left( F_\Lambda
\bar{X}^\Lambda - \bar{F}_\Lambda X^\Lambda \right) \left( -
\tfrac{1}{6} R + D \right) \, . \ee
Upon adding this part to the lagrangian \eqref{1.3} the D-field
equation of motion yields the relation
\be - i \left( F_\Lambda \bar{X}^\Lambda - \bar{F}_\Lambda
X^\Lambda \right) =  2 \cF_{IJ} \left(  v^I  \vb^J + \tfrac{1}{4}
x^I  x^J \right) \, . \ee
Substituting this back into \eqref{1.2} results in an
Einstein-Hilbert term of the form
\be e^{-1} \, \cL =  \, -  R \,  \cF_{IJ} \left(  v^I  \vb^J +
\tfrac{1}{4}  x^I  x^J \right) \, . \ee
In order to have a canonical normalization we choose the
dilatation gauge
\be\label{eq:D-eq} 2 \, \kappa^2 \, \cF_{x^I x^J} \left( v^I \vb^J
+ \tfrac{1}{4} x^I x^J \right) = 1 \, . \ee

Eliminating the fields $\vec{\cV}_{\mu}$ and $D$ together with
imposing the gauge constraints \eqref{eq:SU2-eq} and
\eqref{eq:D-eq} defines the superconformal quotient of the
lagrangian \eqref{1.3}. The resulting theory is a Poincar\'e
supergravity theory coupled to $n_H - 1$ tensor multiplets and one
double tensor multiplet whose scalars parameterize the manifold
$\cM_{\rm TM}$. The corresponding lagrangian reads
\be\label{1.4}
\begin{split}
e^{-1} \cL = & - \tfrac{1}{2 \kappa^2} R + \, \cF_{x^I x^J} \left( \partial_\mu v^I \partial^\mu \vb^J + \tfrac{1}{4} \partial_\mu x^I \partial^\mu x^J - \, E^I_\mu \, E^{J \mu} \right)  \\
&  - \tfrac{1}{2} \vec{S}_\mu M^{-1} \vec{S}^\mu - \tfrac{1}{2}
\vec{T}_\mu M^{-1} \vec{T}^\mu
- \vec{S}_\mu M^{-1} \vec{T}^\mu \\
& \, + \I \, E^I_\mu \left( \cF_{v^I x^J} \, \partial^\mu v^J -
\cF_{\vb^I x^J} \partial^\mu \vb^J \right) \, ,
\end{split}
\ee
where the constraints \eqref{eq:SU2-eq} and \eqref{eq:D-eq} are
implicitly understood and $M^{-1}$ is the inverse of the matrix
\eqref{1.2}. Henceforth we will work in the conventions where
$\kappa^{-2} = 2$.

Taking the lagrangian \eqref{1.4} and dualizing the tensor fields
into scalars by adding a suitable Lagrange multiplier finally
leads to a standard Poincar\'e supergravity theory coupled to
$n_H$ hypermultiplets. The scalars parameterize the manifold
$\cM_{\rm HM}$ which for $n_H=n_V+1$ has precisely the correct
dimension for a manifold in the image of the c-map. Note that the
lagrangian \eqref{1.4} is also completely determined by $\cF(v,
\vb, x)$ which in turn is fixed by the function $H(\eta)$
appearing in the contour integral \eqref{eq:contour}.

\subsection{The classical c-map}
\label{classexample}

Before embarking on the computation of the perturbatively
corrected hypermultiplet moduli space, we need to connect the
classical result \eqref{eq:confc} to the hypermultiplet lagrangian
\eqref{eq:cmap}, using the formalism detailed in the previous
subsection. This computation provides the dictionary between the
variables $x^0, x^\Lambda, v^\Lambda$ appearing on the
superconformal tensor side and the HM scalars coming from the
classical string compactification. In particular this will
identify $x^0$ as the dilaton which then controls the perturbation
series set up in the next section. Our computation thereby
completely parallels the one for the one-loop corrections
presented in Appendix \ref{AppC} from which all intermediate
results may be obtained by setting $c = D(z) = \bar{D}(\bar{z}) =
0$.

\subsubsection{Gauge-fixing the superconformal symmetries}
Our starting point is the function $H(\eta)$ encoding the
classical superconformal c-map \eqref{eq:confc} which is then
substituted into the contour integral \eqref{eq:contour}. To
evaluate this contour explicitly, we perform a partial
gauge-fixing of the $SU(2)/U(1) \subset SU(2)$ symmetries that
belong to the superconformal symmetry group. A convenient
gauge-choice is taken by setting $v^0 = 0$.\footnote{In the
context of the superconformal quotient for hyper-K\"ahler cones
this corresponds to gauge fixing a coordinate on the twistor
space.} The partially gauge fixed function $\cF$ determining the
superconformal TM lagrangian is then given by \cite{RVV}
\be \label{4.0} \cF(x^0, v^\Lambda, \vb^\Lambda, x^\Lambda) = -
{\rm Im}\, \left[ \,\frac{1}{x^0} \oint_{\mathcal C} \frac{\rmd
\zeta}{2 \pi \I \, \zeta} \, F(\eta^\Lambda) \right] \ , \ee
where $F$ is the holomorphic prepotential encoding the VM
couplings of the dual type II compactification and 
${\mathcal C}$ is a positively oriented contour around the
origin. This integral can be evaluated explicitly by using the
homogeneity property $F(\eta) = \frac{1}{\zeta^2} F(\zeta \eta)$
and assuming that $F(\zeta \eta)$ has no poles at the origin of
the complex $\zeta$-plane. This yields
\be \label{4.1} \cF(x^0, v^\Lambda, \vb^\Lambda, x^\La) =
\frac{1}{4 x^0} \left( N_{\La\Si} x^\La x^\Si - 2 K(v, \vb)
\right) \ , \ee
where $K(v, \vb)$ and
$N_{\Lambda \Si}$ are the objects from special geometry defined in
\eqref{2.3}, but now evaluated in terms of the tensor multiplet
scalars $v^\Lambda$.

In the next step we compute the derivatives of $\cF(x^0, v^\La,
\vb^\La, x^\La)$ entering into the lagrangian \eqref{1.3}. For the
derivatives which do not involve the coordinates $v^0, \vb^0$ this
is straightforward. Writing the result in terms of the
inhomogeneous coordinates
\be \label{inhomogeneous}
\begin{split}
A^\La \equiv \frac{x^\La}{2 \, x^0} \ , \qquad z^\La \equiv & \,
\frac{v^\La}{v^1} \, ,
\end{split}
\ee
the resulting expressions are readily be obtained from eqs.\
\eqref{eq:qder} by setting $c = 0$. One can then verify that these
equations satisfy all those constrains in \eqref{eq:Fconst} which do
not involve derivatives with respect to $v^0$. The remaining
conditions cannot be checked since the partially gauge fixed
result \eqref{4.1} does not allow to compute derivatives of
$\cF(x^0, v^\La, \vb^\La, x^\La)$ with respect to $v^0$. This is,
however, not an obstacle when constructing the TM lagrangian as
the derivatives $\cF_{x^0 v^0}, \cF_{x^\La v^0}$ drop out from
\eqref{1.4} due to setting $v^0 = 0 \rightarrow \partial_\mu v^0 =
0$.

We now fix the remaining superconformal gauge symmetries. In order
to break the residual $U(1)$ and dilatation symmetries which are
left after imposing $v^0 = 0$, we set $v^1 = \vb^1$, implementing
the gauge choice \eqref{eq:SU2-eq}. 
Furthermore, we have to solve
the embedding equation \eqref{eq:D-eq} that arises after fixing
the dilatations. Substituting the derivatives of $\cF$ and using
the homogeneity property of the K\"ahler potential \eqref{2.3} the
condition \eqref{eq:D-eq} can easily be solved for $v^1$:
\be \label{v1} v^1 = \sqrt{\frac{4x^0}{K(z,\zb)}} \, . \ee
Here $K(z,\bar z)= z^\La N_{\La\Si} \bar z^\Si$ is understood as a
function of the inhomogeneous coordinates $z, \zb$, eq.\
\eqref{inhomogeneous}. This relation then expresses $v^1 = \vb^1$
in terms of the coordinates $x^0, z^a, \zb^a$. Together with the
gauge condition \eqref{eq:SU2-eq}, eq.\ \eqref{v1} completely
fixes the superconformal symmetries on the tensor side.

\subsubsection{The tensor multiplet lagrangian}
Following the general construction outlined in the previous
subsection we now calculate the inverse of the matrix $M$, eq.\
\eqref{1.2}, and the $SU(2)$-currents $\vec S_\mu$ and $\vec
T_\mu$ given in \eqref{vecS} and \eqref{vecT} taking the gauge
choices \eqref{eq:SU2-eq} and \eqref{v1} into account. Again the
resulting expressions are easily obtained from eqs.\ \eqref{eq:M},
\eqref{eq:ST1} and \eqref{eq:ST2} by setting $\Delta = 0$.
Substituting these results into the lagrangian \eqref{1.4} gives
the TM lagrangian for the classical case
\be \label{classical}
\begin{split}
e^{-1}{\mathcal L}^{\rm TM}_{\rm cl} = & \, - \frac{1}{2(x^0)^2}
(\partial_\mu x^0)^2 + 2 \, {\mathcal G}_{a \bar b}\partial_\mu
z^a \partial^\mu \zb^{\bar b} + \left(\frac{x^0}{2}\right)
({\mathcal N} +\bar{\mathcal N})_{\La\Si}
\partial_\mu A^\La\partial^\mu A^\Si \\
& \, + 2 \, \cT^{\rm cl}_{IJ} \, E_\mu^I E^{J\mu} + i ({\mathcal
N}-\bar{\mathcal N})_{\La\Si}\left[(\partial_\mu
A^\La)E^{\Si\mu}-2(\partial_\mu A^\La)A^\Si E^{0\mu}\right] \, .
\end{split}
\ee
Here ${\mathcal G}_{a \bar b}$ and ${\mathcal N}_{\La\Si}$ are
given in \eqref{eq:KM} and \eqref{eq:cN}, respectively, and the
matrix $\cT^{\rm cl}_{IJ}$ appearing in the tensor field kinetic
term is given by
\be \label{eq:cTcl} \cT^{\rm cl}_{IJ}  = \frac{1}{x^0} \left[
\begin{array}{cc} \frac{1}{x^0} - ({\mathcal N}+\bar {\mathcal
N})_{\La\Si}A^\La A^\Si
& \frac{1}{2} ({\mathcal N}+\bar{\mathcal N})_{\La\Si}A^\La \\
\frac{1}{2}({\mathcal N}+ \bar{\mathcal N})_{\La\Si}A^\Si &
   - \frac{1}{4}({\mathcal N} +\bar{\mathcal
N})_{\La\Si}\end{array}\right] \, . \ee
Upon setting $x^0 = \e^{- \phi}$ this is precisely the classical
tensor multiplet lagrangian \eqref{2.6}. This identification
only makes sense for units in which $\kappa^{-2}=2$. 
One can reinstall $(2\kappa^2)^{-1}$ as an overall factor in front of the 
lagrangian, and in this convention all the fields are dimensionless. 
In particular, our four-dimensional dilaton is dimensionless and
is related to the string coupling constant as
\begin{equation}
{\rm e}^{-\phi_\infty/2}=g_s\ .
\end{equation}
This relation is up to (dimensionless) volume factors of the CY$_3$, 
but we will work in conventions in which we set this to unity.
They are not important for counting powers of $g_s$. The result of our
gauge-fixing condition \eqref{v1} implies that $v^1$ also scales 
like $g_s$. This is consistent with the observation made at the end
of Section 2, where we say that $\eta^1/\eta^0$ scales like $g_s^{-1}$.

\subsubsection{The dual hypermultiplet lagrangian}

Finally, we construct the HM lagrangian dual to \eqref{classical}
by converting the tensor into scalar fields. For this purpose we
add the Lagrange multipliers
\be \label{eq:LMP} e^{-1} \cL^{\rm LM} = 2 \, (\partial_\mu w_0)
\, E^{0\mu} -  (\partial_\mu B_\La) \, E^{\La\mu} \ee
to the TM lagrangian \eqref{classical}. Here the prefactors are
purely conventional and have been chosen for later convenience.
We then eliminate the tensor field strength in favor of the
scalars $w_0$, $B_\La$ by substituting their algebraic equation of
motion back into $\cL^{\rm TM}_{\rm cl} + \cL^{\rm LM}$. This
results in the classical
   HM lagrangian
\be \label{eq:clHM}
\begin{split}
e^{-1}  {\mathcal L}^{\rm HM}_{\rm cl} = & \, - \tfrac{1}{2
(x^0)^2}(\partial_\mu x^0)^2
+ 2 \, {\mathcal G}_{a \bar b}\partial_\mu z^a\partial^\mu \bar z^{\bar b}   \\
& \, + \tfrac{1}{2}  x^0 \, ({\mathcal N}+\bar{\mathcal
N})^{\La\Si}
\left| 2 {\mathcal N}_{\La\Xi} \partial_\mu A^\Xi + i \partial_\mu B_\La\right|^2  - \tfrac{1}{2} (x^0)^2 \left( \partial_\mu w_0 - A^\La\partial_\mu B_\La\right)^2 \, . \\
\end{split}
\ee
Comparing this expression to the one obtained by the classical
Poincar\'e c-map \eqref{eq:cmap} we find complete agreement after
identifying
\be \label{eq:ct}
\begin{split}
x^0 =  {\rm e}^{-\phi}\,, \qquad w_0 =   \sigma + \frac{1}{2}A^\Si
B_\Si \, .
\end{split}
\ee

Furthermore, eq.\ \eqref{eq:ct} allows us to determine the action
of the Peccei-Quinn isometries \eqref{eq:PQ} on the coordinates
$x^0, w_0, A^\La, B_\La, z^\La$ and $\zb^{\La}$:
\be \delta w_0 = \delta \epsilon +  \alpha^\Si \, B_\Si \, ,
\qquad \delta A^\La = \alpha^\La \, , \qquad \delta B_\La=
\beta_\La \, . \ee
By using the definition \eqref{etadef}, and comparing
\eqref{inhomogeneous} to \eqref{eq:ct} the action of the
Peccei-Quinn isometries can then be implemented directly in
superspace. While the shifts associated with $\epsilon$ and
$\beta_\Lambda$ are automatically encoded in the TM description,
$\alpha^\Lambda$ acts non-trivially on the scalars $\eta^I$
\cite{Berkovits:1995cb}:
\be \label{superPQ} \eta^0 \rightarrow \eta^0 \,,~~~~~~~~ \eta^\La \rightarrow 
\eta^\La  + 2\alpha^\La\,\eta^0~. \ee
It is instructive, and will be useful in the next section, to
check directly in superspace that \eqref{4.0} leads to a
lagrangian which is invariant under \eqref{superPQ}. While
\eqref{superPQ} is a gauge-independent statement, we verify this
invariance in the particular gauge $v^0 = 0$. The infinitesimal
variation of \eqref{4.0} under \eqref{superPQ} then gives
\be \label{DF} \delta\left[-{\rm Im}\, \left( \,\frac{1}{x^0}
\oint_{\mathcal C} \frac{\rmd \zeta}{2 \pi \I \, \zeta} \, F(\eta)
\right)\right] = -{\rm Im}\left[2\alpha^\La\, x^\Si
\,F_{\La\Si}(v)\right] \ . \ee
It is straightforward to check that substituting the second
derivatives of this variation into the lagrangian \eqref{1.4}
results in a  total derivative in the action. The latter is then
invariant under the Peccei-Quinn symmetries
\eqref{superPQ}. Equipped with this knowledge we are now
ready to discuss the string loop corrections to the classical
lagrangian \eqref{eq:clHM}.

\section{One-loop corrections}
\label{S4}

In order to identify the perturbatively corrected HM moduli space,
we follow the strategy of \cite{AntoniadisUH} and search for
deformations of the classical result \eqref{eq:clHM} compatible
with string perturbation theory, i.e., satisfying the conditions
(1) - (6) given in the introduction. The superconformal quotient
construction of Section \ref{tensorquotient} shifts the issue
of implementing the QK condition (1) to finding suitable
deformations of $H^{\rm cl}(\eta)$ which obey the conditions (2) -
(6). Our construction proceeds as follows: first we  determine
deformations of $H^{\rm cl}(\eta)$ which are subleading in the
dilaton and preserve the Peccei-Quinn symmetries in Subsection
\ref{SS4.1} thereby implementing the conditions (1) - (4). We then
construct the corresponding HM lagrangian via the superconformal
quotient construction before incorporating the remaining
conditions (5) - (6). This procedure will lead to the
perturbatively corrected HM spaces given at the end of Subsection
\ref{fcc}.

\subsection{Deformations of $H^{\rm cl}(\eta)$}
\label{SS4.1}

We start by investigating deformations of $H^{\rm cl}(\eta)=
\tfrac{F(\eta^{\Lambda})}{\eta^0}$. The QK condition (1) enforces
that such deformations are homogeneous of degree one under a
rescaling of $\eta^I$ and have no explicit $\zeta$ dependence
(cfg.\ Subsection \ref{off-shell}). Next we demand that the
deformations should be subleading in $g_s$, i.e. they are at least 
of order $g_s^2$ higher than the classical result. 
The powers of $g_s$ can be counted
using the fact that $\eta^1/\eta^0$ scales like $g_s^{-1}$. Moreover,
in the gauge $v^0 = 0$, $\eta^0=x^0={\rm e}^{-\phi}$ is of order $g_s^2$.
From this one can see that for the universal hypermultiplet, the first
term in \eqref{qcuhm} is of order $g_s^0$ while the second is of order
$g_s^2$, which is the correct counting for the one-loop correction.
These considerations naturally suggests the following candidates
for correction terms to $H^{\rm cl}(\eta)$:\footnote{Here and in
the following we will not consider terms containing higher powers
of $\ln(\eta^0)$. Such terms typically violate the homogeneity
properties. Moreover, they give rise to terms polynomial in $\phi
= - \ln({\rm e}^{-\phi})$ in the effective action, which do not
occur in string perturbation theory. Deformations containing
higher powers of $\eta^0$ will be discussed in Section \ref{NRT}.
Deformations with lower powers (i.e. negative powers) in $\eta^0$ 
are not consistent with the Peccei-Quinn isometries \eqref{superPQ}.}
\bea\label{eq:def1} H^{\rm def}_1(\eta) & = & D_1(\eta^{\La}) \, ,
\\ \label{eq:def2} H^{\rm def}_2(\eta) & = & D_2(\eta^{\La}) \,
\ln(\eta^0) \, , \\ \label{eq:def3} H^{\rm def}_3(\eta) & = &
D_3(\eta^{\La}) \, \eta^0 \, , \\ \label{eq:def4} H^{\rm
def}_4(\eta) & = & D_4(\eta^{\La}) \, \eta^0 \, \ln(\eta^0) \, .
\eea
Here the functions $D_1(\eta^{\La})$ and $D_2(\eta^{\La})$ are
homogeneous of degree one while $D_3(\eta^{\La})$ and
$D_4(\eta^{\La})$ are homogeneous of degree zero.

Let us now discuss how these deformations contribute to the
superconformal TM lagrangian \eqref{1.4}. In this course we
compute their contribution to the function $\cF(x,v,\vb)$ by
evaluating the contour integral \eqref{eq:contour} for the same
gauge and contour as in the classical case. Substituting the
second derivatives of the resulting functions $\cF(x,v,\vb)$ into
the TM lagrangian then shows that the deformations \eqref{eq:def1}
and \eqref{eq:def3} give rise to surface terms only. Furthermore,
a careful analysis of the contributions arising from the
deformation \eqref{eq:def2} yields that the corresponding terms
contain an odd number of RR fields. Thus \eqref{eq:def2} violates
condition (4) and \emph{is not compatible} with string
perturbation theory.

Hence we are left with considering the deformation
\eqref{eq:def4}. As it will turn out in the subsequent subsections
this deformation indeed satisfies the remaining conditions
(3)-(6). Denoting $D_4(v) = -4 c D(v)$, where the numerical factor
$-4c$ has been extracted for later convenience, we thus make the
following ansatz for the ``loop-corrected'' function $H^{\rm
qc}(\eta^I)$:
\be \label{ansatz} H^{\rm qc}(\eta^I)=\frac{F(\eta^\La)}{\eta^0}-4
\, c \, \tF(\eta^\La) \, \eta^0 \, \ln \,\eta^0 \, . \ee

We then proceed to the evaluation of the contour integral
\eqref{eq:contour} with $H^{\rm qc}(\eta)$. As in the classical
case, the integral is
   carried out in the gauge $v^0=\bar v^0=0$ and with
the contour $\cC$ taken around the origin. Assuming that $D(\zeta
\eta^\La)$ has no poles at $\zeta = 0$ the computation is
straightforward and yields
\be \label{result} {\mathcal F}(x^0,v^\La,\vb^\La,x^\La) =
\frac{1}{4 x^0} \left( N_{\La\Si} x^\La x^\Si - 2 K(v,\vb) \right)
- 2ic \left[{\tF}(v^\La)-\bar{\tF}(\bar v^
\La)\right]\,x^0\,\ln\,(x^0) \, . \ee

At this stage we are just left with verifying condition (3),
namely that \eqref{ansatz} leads to an action which is
invariant under the Peccei-Quinn transformations \eqref{superPQ}. Since
the classical piece already respects these symmetries (cfg.\
\eqref{DF}), we need to check the deformation piece only. Under
the variation \eqref{superPQ}, eq.\ \eqref{eq:def4} becomes
\be \label{contourPQ} \delta {\mathcal F}_4^{\rm def}(v,\bar v,x)
= 
8\,c\,\alpha^\La {\rm Im} \oint_{\mathcal C}\frac{\rmd\zeta}{2\pi i}
\left[D_\La(\zeta \eta^\Si)(\eta^0)^2\,\ln\,(\eta^0)\right] \, .
\ee
This vanishes as long as $D_\La(\zeta\eta)$ has no poles around
the origin. With this proviso, \eqref{ansatz} will lead to
quaternion-K\"ahler metrics invariant under the perturbative
Peccei-Quinn symmetries.

Let us end this subsection with a remark about the universal
hypermultiplet. It was found in \cite{UHMsuper} that the
one-loop corrected HM lagrangian \cite{AntoniadisUH} can be
encoded in $H^{\rm UHM}(\eta)$, eq.\ \eqref{qcuhm}. We observe
that the ansatz \eqref{ansatz} naturally reduces to this equation
if we set $D(v) = -i$. Up to a rescaling by a constant (which can
be absorbed in $c$) this is, however, the only non-trivial $D(v)$
allowed in the absence of non-universal hypermultiplets. Thus our
ansatz naturally generalizes the result for the universal
hypermultiplet.

\subsection{The deformed tensor multiplet lagrangian}
\label{SS:5.2}

We now proceed by computing the perturbative corrections to the
classical TM lagrangian which arise from \eqref{result}. Since
this calculation is lengthy and rather technical its details are
given in Appendix \ref{AppC}. Here we just quote the final result
for the deformed TM lagrangian. With the same conventions as in
\eqref{inhomogeneous} and \eqref{eq:ct}, we have
\be \label{onelooptensorlagrangian}
\begin{split}
e^{-1}{\mathcal L}_{\rm qc}^{\rm TM}= & \,-\frac{1+2\dd {\rm
e}^{-\phi}}{2(1+\dd {\rm e}^{-\phi})}(\partial_\mu \phi)^2 +
2(1+\dd {\rm e}^{-\phi}) {\mathcal
G}_{a \bar b}\partial_\mu z^a\partial^\mu \bar z^{\bar b}+ \frac{c^2\, {\rm e}^{-2\phi}}{8(1+\dd {\rm e}^{-\phi})}\tilde A_\mu\tilde A^\mu \\
& \, + \frac{{\rm e}^{-\phi}}{2}({\mathcal M}+\bar{\mathcal
M})_{\La\Si}\partial_\mu A^\La\partial^\mu A^\Si + 2 {\mathcal
T}^{\rm qc}_{IJ}E_\mu^I E^{J\mu} \\
& \, + i \left({\mathcal M}-\bar{\mathcal M}\right)_{\La\Si}
\left[E_\mu^\La\partial^\mu A^\Si - 2 E_\mu^0 A^\La\partial^\mu
A^\Si\right] \\
& \, - 2\dd A_\mu E^{0 \mu} + 2c \left[D(z)+\bar D(\bar
z)\right](\partial_\mu \phi)E^{0\mu} ~.
\end{split}
\ee
In this expression, we have introduced the notation 
\be\label{eq:dd}
\dd = \frac{ic}{2} \left[\tF(z)-\bar\tF(\bar z)\right], 
\ee
$\cG_{a \bar{b}}(z,\bar z)$ is the local special K\"ahler metric
\eqref{eq:KM} arising from the prepotential of the dual VM
geometry, and the matrix $\cM_{\Lambda \Sigma}$ appearing in the
kinetic term of the RR scalars is given by
\begin{equation}\label{eq:def}
{\cal M}_{\Lambda\Sigma}(z,\bar z,\phi)=-i{\bar
F}_{\Lambda\Sigma}-\frac{(Nz)_\Lambda
(Nz)_\Sigma}{(zNz)}\Big(1+{\rm e}^{-2\phi}\frac{\Delta^2}{|{\tilde
M}|}\Big)-\frac{\Delta\,|v^1|^2}{4|\tilde M|}(Nz)_\Lambda (N{\bar
z})_\Sigma\ .
\end{equation}
The factor of $|v^1|^2$ appears as a result of our gauge fixing
procedure. Similar to \eqref{v1}, we set $v^1$ to be real and
solve \eqref{eq:D-eq} at the one-loop level,
\be \label{gf-oneloop} v^1 = \sqrt{\frac{4x^0(1+\dd x^0)}{K(z,\bar
z)}} \, .\ee
The quantity $|\tilde M|$ is related to the determinant of the
matrix $M$ appearing in the tensor multiplet lagrangian
\eqref{1.4}. Applied to our situation, it is given by
\begin{equation}\label{mtilde}
|\tilde M|=\frac{{\rm e}^{2\phi}|v^1|^4}{16}|z^\Lambda
N_{\Lambda\Sigma} z^\Sigma|^2-{\rm e}^{-2\phi}\Delta^2\ .
\end{equation}
It is clear that in the classical limit, $\Delta=0$, the matrix
${\cal M}$ coincides with ${\cal N}$ defined in \eqref{eq:cN}.
Furthermore, the kinetic term for the tensor fields is determined
through
\be
\begin{split}
{\mathcal T}^{\rm qc}_{IJ} = &\, \left[\begin{array}{cc} f{\rm
e}^{2\phi}+ 4{\mathcal T}^{\rm qc}_{\La\Si}A^\La A^\Si & -2
{\mathcal T}^{\rm qc}_{\Si\Xi} A^\Xi \\
-2 {\mathcal T}^{\rm qc}_{\La\Xi} A^\Xi & {\mathcal T}^{\rm qc}
_{\La\Si}
\end{array}\right]~,~~~~~~~f=\frac{1+2\dd\,{\rm e}^{-\phi}}{1+\dd\,{\rm
e}^{-\phi}}\ ,
\end{split}
\ee where
\begin{equation} \label{eq:cTqc}
-4{\rm e}^{-\phi}{\mathcal T}_{\Lambda\Sigma}^{\rm qc}=-i{\bar
F}_{\Lambda\Sigma}-\frac{(Nz)_\Lambda
(Nz)_\Sigma}{(zNz)}\Big(1+{\rm e}^{-2\phi}\frac{\Delta^2}{|{\tilde
M}|}\Big)+\frac{\Delta\,|v^1|^2}{4|\tilde M|}(Nz)_\Lambda (N{\bar
z})_\Sigma + {\rm h.c.}\ .
\end{equation}
Notice the similarity with \eqref{2.7} to which it reduces in the
classical limit. Finally, $A_\mu$ and $\tilde A_\mu$ are given by
the K\"ahler connection
\be \label{4.12} A_\mu =  \frac{i}{K(z,\bar z)} \left[(N\bar z)_a
\partial_\mu z^{ a} - (Nz)_{\bar a}
\partial_\mu \bar z^{\bar a} \right] \, , \ee
and
\be \label{4.11}
\begin{split}
\tilde A_\mu = & \, i\left[\partial_a \tF (z)\partial_\mu z^a -
\partial_{\bar a}\bar\tF (\bar z)\partial_\mu\bar z^{\bar a}\right] \, ,
\end{split}
\ee
respectively.

Let us now comment about the structure of the one-loop corrected
lagrangian \eqref{onelooptensorlagrangian} and contrast it with
the classical expression \eqref{classical}. We first of all remark
that the dilaton kinetic term is modified. Comparing to the
universal hypermultiplet \eqref{A.42} this term has the same
functional dependence on the dilaton, but in the generic case the
constant $c$ can be promoted to a function of the geometric moduli
$z, \zb$. The terms involving the squares of the RR scalars and
the tensors have the same structure as in the classical case, but
the classical couplings ${\mathcal N}$ and ${\mathcal T}^{\rm cl}$
are now replaced by their ``quantum corrected'' counterparts
${\mathcal M}$ and ${\mathcal T}^{\rm qc}$. However, due to the
different sign in the last term, it is not the combination
${\mathcal M}+\bar{\mathcal M}$ which enters into ${\mathcal
T}^{\rm qc}$, as one might have expected from \eqref{eq:cTcl}. In
fact this sign difference is necessary for obtaining the quantum corrected
universal hypermultiplet in Appendix \ref{SB.4}. 

The metric for the geometric moduli $z, \zb$ receives two kinds of
loop-corrections: the first is the fiberwise rescaling by $(1+\dd
{\rm e^{-\phi}})$ along the dilaton direction (possibly with
$z$-dependent corrections to the metric, encoded in $\dd$). The
second kind of corrections (third term in the first line of
\eqref{onelooptensorlagrangian}) depends quadratically on $\tilde
A_\mu$ and induces explicit non-K\"ahler $\partial z\partial z$,
$\partial\bar z\partial\bar z$ terms, in addition to further
correcting the mixed terms. These terms disappear however when
$D(z)$ is constant. Finally, there are also the quantum mixing
terms in the last line of \eqref{onelooptensorlagrangian}.

Let us end this subsection by noting that upon setting $c=\dd =0$
the quantum corrected TM lagrangian
\eqref{onelooptensorlagrangian} reduces to the classical result
\eqref{classical}, consistently with turning off the deformation
in \eqref{result}. See Appendix \ref{SB.2} for details.
\subsection{The deformed hypermultiplet lagrangian}
With the result \eqref{onelooptensorlagrangian} at hand we now
compute the corresponding
  HM lagrangian by dualizing the tensors $E_\mu^0$, $E^\Lambda_\mu$ to scalars $w_0$,
$B_\Lambda$. Again, the details of the dualization can be found in
Appendix \ref{AppC}. The final result for the deformed HM
lagrangian then reads
\be \label{finallagrangian}
\begin{split}
e^{-1}{\mathcal L}_{\rm qc}^{\rm HM} = & \, -\frac{1+2 \dd\,{\rm
e}^{-\phi}}{2(1+\dd\,{\rm e}^{-\phi})}(\partial_\mu \phi )^2 +
2(1+\dd {\rm e}^{-\phi}){\mathcal
G}_{a \bar b}\partial_\mu z^a\partial^\mu \bar z^{\bar b} +\frac{c^2\,{\rm e}^{-2\phi}}{8(1+\dd {\rm e}^{-\phi})}\tilde A_\mu\tilde A^\mu\\
& \, - \frac{{\rm e}^{-2\phi}(1+\dd\,{\rm
e}^{-\phi})}{2(1+2\dd\,{\rm e}^{-\phi})} \left|\partial_\mu w_0 -
\dd A_\mu +{c}\left[D(z)+\bar D(\bar z)\right](\partial_\mu \phi)
   - A^\La\partial_\mu B_\La\right|^2 \\
& \, - \frac{1}{8} (\cT^{\rm qc})^{\La\Si} \left( 2 {\mathcal
M}_{\La\Xi}
\partial^\mu A^\Xi + i \partial^\mu
   B_\La\right)\left( 2 \bar{\mathcal M}_{\Si\Xi} \partial_\mu
   A^\Xi
- i \partial_\mu B_\Si \right) \\
& \, - {\rm Re}\left[\dd {\rm e}^{-\phi} \frac{(1+\dd {\rm
e}^{-\phi} )}{2|\tilde M| K} (Nz)_\La (N\bar z)_\Xi (\cT^{\rm
qc})^{\Xi\Pi} ({\mathcal M}+\bar{\mathcal M})_{\Pi\Si}\partial_\mu
A^\La
\partial^\mu A^\Si\right]~.
\end{split}
\ee
Here $(\cT^{\rm qc})^{\La \Si}$ is the inverse of \eqref{eq:cTqc}.
An alternative way of writing this lagrangian is given in \eqref{eq:B30}.

One should emphasize that this family of metrics is QK {\it by
construction}.\footnote{As a non-trivial consistency check, we
used Mathematica to explicitly verify the Einstein property of
these metrics in several examples including two hypermultiplets.}
Furthermore, it is completely specified by \emph{two} holomorphic
functions, the prepotential $F(v)$ (homogeneous of degree two)
determining the classical result and $cD(v)$ (homogeneous of
degree zero) encoding the ``quantum deformations''. In order to
establish that these quantum deformations indeed describe string
loop corrections to the classical HM moduli space, we still have
to require that the metrics \eqref{finallagrangian} satisfy the
conditions (5)-(6) from the introduction. We will then implement
the requirement (5) here, and leave (6) for the next
subsection.

Concerning condition (5), we use that parity acts on the
fields appearing in \eqref{finallagrangian} by
\be \label{4.15} w_0\leftrightarrow -w_0\,,~~~~~B_\La
\leftrightarrow - B_\La\,,~~~~~A^\La\leftrightarrow A^\La\,,~~~~~z^\La\leftrightarrow \bar z^\La \,
. \ee
Applying that $F_{\La\Si}(z)=-\bar F_{\La\Si}(z)$, this transformation is
a symmetry of the lagrangian if
\be D(z) = -\bar D(z)~. \ee
Thus invariance under parity imposes the restriction that $D(z)$
is purely imaginary. We further remark that the RR scalars
$A^\La$ and $B_\La$ enter into \eqref{finallagrangian} in pairs
only so that (4) is satisfied automatically.

\subsection{Matching to string loop amplitudes}
\label{fcc}
Finally, let us compare the lagrangian
\eqref{finallagrangian} to known results on string loop-corrected
HM metrics, implementing condition (6). The only undetermined quantities
of our lagrangian are a numerical constant $c$, and a holomorphic function
$D(z)$. We will fix them by comparing to known results for string loop 
amplitudes.

We start with the sector of the universal hypermultiplet. This can be 
obtained from compactifications of IIA strings on a rigid ($h^{1,2}=0$)
CY$_3$, so there are no geometric moduli $z^a$. 
The one-loop correction was determined in \cite{AntoniadisUH} 
and it suffices to look at the kinetic term of the dilaton only. 
In Einstein frame, the one-loop correction can be written as
\begin{equation}
e^{-1}{\cal L}^{{\rm UHM}}=-\frac{1}{2}\,\frac{(1-2\chi_1{\rm e}^{-\phi})}{(1-\chi_1
{\rm e}^{-\phi})}\,(\partial_\mu \phi)^2+\cdots \ .
\end{equation}
Here, $\phi$ is related to the four-dimensional dilaton in a way explained in 
\cite{AntoniadisUH}, and the one-loop constant 
$\chi_1$ is proportional to the Euler number $\chi=2(h^{1,1}-h^{1,2})$ of 
the CY$_3$. The numerical value of $\chi_1$ was determined in 
\cite{AntoniadisUH}, generalizing
previous results in \cite{Kiritsis:1994ta},
\begin{equation}
\chi_1 = \frac{4\zeta(2)\chi}{(2\pi)^3} = \frac{1}{6\pi}(h^{1,1}-h^{1,2})\ .
\end{equation}
Comparing now to our general effective action \eqref{finallagrangian}, 
we determine that
\begin{equation}
c=-\chi_1=-\frac{\chi}{12\pi}\ .
\end{equation}
Here we have used that the function $D$ in \eqref{eq:dd} must be an 
imaginary constant, since there are no geometric moduli for rigid Calabi-Yau's.
We have normalized it such that $D=-i$.

We now look at the generic compactifications, 
and consider the cases of IIA and IIB 
separately. Because we have a common description for universal and 
non-universal hypermultiplets, the value of $c$ is fixed and 
also appears in the one-loop corrections to the kinetic terms of the 
geometric moduli $z^a$. This sector has been considered in 
\cite{HM1}, and in Einstein frame these corrections
were found to take the form
\be \label{antoniadisbisIIa} S_{\rm IIA} =
\frac{1}{2\kappa^2_4}\int {\rm d}^4x \sqrt{- g^{(E)}}\,2\left(1+\chi_2 {\rm
e}^{-\phi}\right){\mathcal G}^{0}_{a \bar b
}\partial_\mu z^a \partial^\mu \bar z^{\bar b}~, 
\ee
for type IIA, and %
\be \label{antoniadisbisIIb} S_{\rm IIB} =
\frac{1}{2\kappa^2_4} \int {\rm d}^4x \sqrt{- g^{(E)}}
\,2\left(\cG_{a \bar{b}} - \chi_2{\rm e}^{-\phi}
\left( {\mathcal G}^0_{a \bar b
} + \ldots \right) \right)\partial_\mu z^a \partial^\mu \bar z^{\bar b}~, \ee
for type IIB. Here, $\chi_2$ is a parameter that characterizes the one-loop
correction. Furthermore, the special K\"ahler metrics $\cG_{a \bar{b}}$ and $\cG_{a \bar{b}}^0$ are computed from the prepotential\footnote{On top of the perturbative corrections, $F(X)$ receives contributions from worldsheet instantons,
but we refrain from giving explicit expressions.}
\begin{equation}\label{prepot}
F(X)=\frac{1}{3!}\kappa_{abc}\frac{X^aX^bX^c}{X^1}-i
\frac{\zeta(3)\chi}{2(2\pi)^3}(X^1)^2  \ , 
\ee
where $\cG_{a \bar{b}}$ arises from the full prepotential (including perturbative $\alpha^\prime$ corrections encoded in the second term) while $\cG_{a \bar{b}}^0$ comes from the first term only. This implies that, at tree level in $g_s$, the $\alpha^\prime$ corrections occurring in the type IIB case are encoded in $\cG_{a \bar{b}}$ while possible $\alpha^\prime$ corrections at one-loop are contained in the dots. Notice that these corrections are absent in the type IIA case. Additionally it was shown that eqs.\ \eqref{antoniadisbisIIa} and \eqref{antoniadisbisIIb} can be understood from the dimensional reduction of the ten-dimensional effective
actions on a CY$_3$ if the $R^4$ terms are taken into account. Finally, we note that the one-loop corrections in type IIA and type IIB come with a different sign.
This is consistent with mirror symmetry (now at one-loop
in $g_s$) which sends $\chi \leftrightarrow -\chi$ and exchanges
IIA $\leftrightarrow$ IIB.  

After these preliminaries, we are now ready to determine the
one-loop deformation $cD(z)$ by comparing the lagrangian
\eqref{finallagrangian} to \eqref{antoniadisbisIIa} and
\eqref{antoniadisbisIIb} in the type IIA and IIB case,
respectively. Considering the type IIA case 
we observe that, consistently
with four-dimensional ${\mathcal N}=2$ supersymmetry, the
corrections to the HM metric arise entirely  in $g_s$. Comparing the appropriate
subsectors of \eqref{finallagrangian} and \eqref{antoniadisbisIIa}
leads to $\Delta = \chi_2$. Recalling eq.\ \eqref{eq:dd} this amounts to having $D(z)$ equal
to (a purely imaginary) constant, which we normalize to be $-i$.
We then have
\be \label{identff} D_{\rm IIA}= -i~,~~~~~~~c=\chi_2.
\ee
But from the universal sector, we had already concluded that $c=-\chi_1$.
Therefore, $\chi_2$ is fixed by supersymmetry to be $\chi_2=-\chi_1$.
The value of $\chi_2$ found in \cite{HM1} differs from ours by a factor of -2. 
It would be interesting to resolve this apparent mismatch.

The IIB case is more complicated due to the presence of $\alpha'$-corrections. First observe that $\dd = (ic/2) \left[ D(z) - \bar{D}(\bar{z}) \right]$ is the sum of a holomorphic and anti-holomorphic function. Clearly, it is implausible that volume dependent terms, with
\be 
{\mathcal V}= -\frac{i}{6}\kappa_{abc} (z^a -\bar z^a)(z^b
-\bar z^b)(z^c-\bar z^c)~, 
\ee
separate into the sum of holomorphic and anti-holomorphic expressions in such a way that they can contribute to $\dd$. Comparing to the expression \eqref{finallagrangian} this implies that the dots appearing in \eqref{antoniadisbisIIa} complete $\cG^0_{a \bar{b}}$ into the full $\alpha^\prime$-corrected metric $\cG_{a \bar{b}}$. As a consequence \eqref{antoniadisbisIIa} becomes
\be \label{eq:4.25} S_{\rm IIB} =
\frac{1}{2\kappa^2_4}\int {\rm d}^4x \sqrt{- g^{(E)}}\left(1 -
{\rm e}^{-\phi}\chi_2\right){\mathcal G}_{a \bar b
}\partial_\mu z^a \partial^\mu \bar z^{\bar b}~. \ee
{}From this we then read off
  \be \label{identffb} D_{\rm IIB}=
i~,~~~~~~~c=\chi_2~. \ee

A constant $D(z)$ leads to further simplifications of the
expression \eqref{finallagrangian}. Thus we can state our final
result for the type IIA compactification as (for IIB, change $c$
into $-c$):

\be \label{finalfinallagrangian}
\begin{split}
e^{-1}{\mathcal L}_{\rm qc}^{\rm HM} = & \, -\frac{1+2c \,{\rm
e}^{-\phi}}{2(1 + c\,{\rm e}^{-\phi})}(\partial_\mu \phi )^2 + 2(1
+ {c}\, {\rm e}^{-\phi}){\mathcal
G}_{a \bar b}\partial_\mu z^a\partial^\mu \bar z^{\bar b} \\
& \, - \frac{{\rm e}^{-2\phi}(1 + {c} \,{\rm e}^{-\phi})}{2(1+ 2
{c}
\,{\rm e}^{-\phi})} \left|\partial_\mu w_0 - c A_\mu - A^\La\partial_\mu B_\La\right|^2 \\
& \, - \frac{1}{8} ({\mathcal T}^{\rm qc})^{\La\Si} \left( 2
{\mathcal M}_{\La\Xi}
\partial^\mu A^\Xi + i \partial^\mu
   B_\La\right)\left( 2 \bar{\mathcal M}_{\Si\Xi} \partial_\mu
   A^\Xi
- i \partial_\mu B_\Si \right) \\
& \, - {\rm Re}\left[{ c}\, {\rm e}^{-\phi} \frac{(1 + {c}\, {\rm
e}^{-\phi} )}{2|\tilde M| K} (N z)_\La (N\bar z)_\Xi ({\mathcal
T}^{\rm qc})^{\Xi\Upsilon} ({\mathcal M}+\bar{\mathcal
M})_{\Upsilon\Si}\partial_\mu A^\La
\partial^\mu A^\Si\right]~.
\end{split}
\ee
We remark that, for the type IIA case, the appearance of the 
K\"ahler connection in the second line of \eqref{finalfinallagrangian} 
was already anticipated in \cite{Gunther}, based on an analogy to 
heterotic strings.

\section{Higher loop corrections} \label{NRT} 
A natural question is to ask about the presence of higher loop
corrections. For the case of the universal hypermultiplet only,
this was analyzed in \cite{AntoniadisUH}. There it was concluded
that higher loop corrections are present, but can in fact be
absorbed into a field redefinition, or equivalently, by a
coordinate transformation on the HM moduli space. In other words,
if the appropriate variables are chosen, there are no additional
corrections beyond one-loop. We will argue in this section that
such a non-renormalization theorem might also work in the generic
case including an arbitrary number of hypermultiplets.

Non-renormalization theorems are best understood in an off-shell
formulation. We have seen that at one-loop order, the superspace
lagrangian is given by \eqref{ansatz} and that $\eta^0$ is the
dilaton multiplet, at least in the chosen gauge $v^0=\bar v^0=0$
for which $x^0={\rm e}^{-\phi}= g_s^2$. In this gauge, there is a
natural generalization to higher loop corrections,
\be \label{higherloops} {\cal F}(v,\bar v,x) =-{\rm
Im}\sum_{n=0}^{\infty}\left[\oint_{{\cal C}_0}\frac{\rmd
\zeta}{2\pi i \zeta}(x^0)^{n-1}F_{n}(\eta^\Lambda)\right]\ , \ee
where $F_0(\eta^\Lambda)$ and $F_2(\eta^\Lambda)$ are given in
\eqref{ansatz} and the contour is again chosen around the origin.
Furthermore, as explained in the beginning of the previous
section, we have that $F_1=0$. The $F_n$ with $n\geq 3$ are
coefficient functions defining possible higher loop corrections.
The homogeneity condition needed for superconformal invariance
requires $F_n$ to be homogeneous of degree $2-n$. If needed,
logarithms of $x^0$ could be included in $F_n$, as is the case
e.g. for $n=2$, as long as the homogeneity conditions are
satisfied under the contour integral.

It is now easy to see that all terms with $n\geq 3$ vanish under
the contour integral. Indeed, using the homogeneity properties of
the $F_n$, \eqref{higherloops} can be rewritten as
\be {\cal F}(v,\bar v,x) = -{\rm
Im}\sum_{n=0}^{\infty}\left[(x^0)^{n-1}\oint\frac{\rmd \zeta}{2\pi
i}\,\zeta^{n-3}F_{n}(\zeta \eta^\Lambda)\right]\ , \ee
with, as usual, $\zeta \eta^\Lambda=v^\Lambda+x^\Lambda\zeta-{\bar
v}^\Lambda \zeta^2$, and the contour is chosen around the origin.
If we now assume that the $F_n(\zeta \eta^\Lambda)$ have no
singularities around the origin in the $\zeta$ plane, then it is
clear that all terms with $n\geq 3$ vanish under the contour
integral. This amounts to a non-renormalization theorem for
hypermultiplets (or, equivalently, off-shell tensor multiplets).

The following important remark is in order. Notice that we have
assumed the contour to be around the origin. This choice is
intimately related to the choice of $SU(2)$ gauge, and the
identification $x^0={\rm e}^{-\phi}$. We could relax this
assumption. If different contours are taken, it might be that they
enclose poles of $F_n$ away from the origin, which could yield
non-vanishing contributions to the contour integral. In fact, one
should consider the more $SU(2)$ covariant expression
\be \label{hl-cov} {\cal F}(v,\bar v,x)=-{\rm
Im}\sum_{n=0}^{\infty} \left[\oint_{\cal C}\frac{\rmd \zeta}{2\pi
i \zeta}(\eta^0)^{n-1}F_{n}(\eta^\Lambda)\right]\ , \ee
where the contour encloses all poles. These would yield
non-vanishing higher-loop corrections. In fact, we know from
\cite{HM1} and \cite{AntoniadisUH} that such corrections can
appear. It might however be that these higher-loop corrections can
be absorbed by redefining the dilaton. In our language, this would
correspond to a modification of the identification between $x^0$
and ${\rm e}^{-\phi}$.

In the gauge $v^0=0$, \eqref{hl-cov} reduces to
\eqref{higherloops}, in which the absense of higher-loop
corrections is manifest, provided the contour is taken to be
around the origin. This choice of contour works for sure at
one-loop order. To fully prove the non-renormalization conjecture,
a better understanding of how to choose the contours would be
needed. We leave this for future investigation.

\section{Discussion and Outlook}
In this paper we combined superspace and conformal calculus
techniques to determine the string loop corrections to the
hypermultiplet moduli space of type II string theory compactified
on a generic Calabi-Yau threefold. The resulting lagrangian is
completely specified by a single function $H(\eta^I)$ homogeneous
of degree one. The result of our analysis is that this function is
given by
\be\label{7.1} H^{\rm qc}(\eta^I) = \frac{1}{\eta^0} \,
F(\eta^\Lambda) + \frac{\chi}{3\pi} \, D(\eta^\Lambda)\,\eta^0
\ln(\eta^0) \ . \ee
The $\eta$-symbols are $n_H+1$ ${\cal N}=2$ tensor multiplets
which include the dilaton multiplet. The first term corresponds to
the classical term and was obtained in \cite{RVV}. Here
$F(\eta^\Lambda)$ is homogeneous of degree two, and arises from
the vector multiplets after doing the c-map. The second term in
\eqref{7.1} is proportional to the Euler number of the Calabi-Yau
$\chi$ and describes the one-loop corrections. It contains a
function $D(\eta)$, homogeneous of degree zero. To match with
known type II string amplitudes  this function should be taken constant:
\be
D_{\rm IIA} = - i \; , \qquad D_{\rm IIB} = i \, .
\ee
We furthermore argued that there is a
non-renormalization theorem, excluding possible higher loop
corrections. It is fair to say that this is a conjecture rather
than a theorem.

It is somewhat intriguing that our generic effective action allows
for a non-constant holomorphic function $D(z)$, homogeneous of degree zero. At
present, no string amplitude at one-loop seems to contribute to
$D(z)$ apart from the constant term. It would be interesting to
know if this is an exact result to all orders in $\alpha'$.

It remains an open problem to determine the non-perturbative
corrections to the hypermultiplet moduli space which arise from
Euclidean $D$-branes wrapping supersymmetric cycles in the
Calabi-Yau threefold \cite{BBS}. These non-perturbative
corrections are expected to break the Peccei-Quinn symmetries,
making a description in terms of tensor multiplets more difficult,
unless one takes into account global issues in the scalar-tensor
duality (see e.g. \cite{Theis:2002er,Davidse:2004gg}).

A less ambitious extension of our present work is to consider the
gauging of the isometry group of our perturbatively corrected
hypermultiplet moduli spaces\footnote{For the perturbatively
corrected universal hypermultiplet such an investigation was
performed in \cite{Behrndt:2003bf,Membrane}.} along the lines
\cite{italians} which corresponds to considering Calabi-Yau
compactifications with non-trivial background fluxes. Based on the
observations in orientifold compactifications of the type II
string to $\cN = 1$ supergravity \cite{N=1stuff,N=1inst} one would
expect that perturbative corrections can play an important role in
altering the vacuum structure and stabilizing the moduli of the
compactification. In this context it would also be interesting to
consider the orientifold projection of our string loop corrected
$\cN = 2$ supergravity along the lines
\cite{mci,GrimmLouis,orientifold1}. This could lead to results
complementary to the ones obtained in \cite{BHK} where
perturbative corrections to the K\"ahler potential describing the
vector multiplet sector have been studied.

Finally let us remark that based on the string-string duality
\cite{hetexample} between the type IIA string compactified on
Calabi-Yau threefolds and the heterotic string on $K3 \times T^2$,
we expect that our results are also valid for heterotic string
compactifications. In this context it would certainly be
interesting to understand how the string loop corrected
hypermultiplet moduli spaces found in this paper arise in the dual
heterotic picture, since there the corresponding moduli space is
exact in  the string coupling constant.

\bigskip
{\large\bfseries Acknowledgements} \smallskip

\noindent We are grateful to I.\ Antoniadis for his comments on an earlier
draft of this paper. We also like to thank B.\ de Wit, E.\ Imeroni, A.\
Kashani-Poor, B.\ K\"ors, and M. Ro\v{c}ek for stimulating
discussions. S.V. thanks the Weizmann Institute in Rehovot for
hospitality during the final stages of this project. F.S.\ is
supported by a European Commission Marie Curie Postdoctoral
Fellowship under contract number MEIF-CT-2005-023966. D.R.\ is partly 
supported by the European Union RTN network MRTN-CT-2004-005104.
We also acknowledge Intas grant NWO-project 047017015 for support.

\bigskip

\appendix
\section{Derivation of the loop-corrected lagrangians}
\label{AppC}

This appendix collects the technical details for deriving the
quantum-corrected hyper- and tensor multiplet lagrangians
presented in Section \ref{S4}. The calculation follows the
template outlined in Subsection \ref{TMscq} and naturally splits
into the three parts of determining the superconformal gauge
fixing conditions, constructing the tensor multiplet lagrangian,
and dualizing the tensor into hypermultiplets. These steps will be
carried out in Subsections \ref{SB.1}, \ref{SB.2} and \ref{SB.3},
respectively. As an example, we will discuss the quantum corrected
universal hypermultiplet in Subsection \ref{SB.4}. Note that the
construction of the classical results given in Subsection
\ref{classexample} is completely analogous to the quantum case, so
that its intermediate steps can simply be obtained from the
quantum formulae by switching off the deformation piece by setting
$c = D(v) = \Delta = 0$.

\subsection{Gauge fixing the superconformal symmetries}
\label{SB.1}
%
The starting point of our construction is the function
\eqref{result}
\be {\mathcal F}(x^0,v^\La,\vb^\La,x^\La) = \frac{1}{4 x^0} \left(
N_{\La\Si} x^\La x^\Si - 2 K(v, \vb) \right) - 2ic
\left[{\tF}(v^\La)-\bar{\tF}(\bar v^ \La)\right]\,x^0\,\ln\,(x^0)
\, , \ee
which is obtained by evaluating the contour integral of $H^{\rm
qc}(\eta)$, eq.\ \eqref{ansatz}. We then compute the second
derivatives of $\cF$ which enter into the gauge-fixed lagrangian
\eqref{1.4}. Adopting the $SU(2)$ gauge-choice \eqref{eq:SU2-eq}
where $v^0 = 0, v^1 = \vb^1$ and in terms of the inhomogeneous
coordinates \eqref{inhomogeneous} these derivatives read
\be\label{eq:qder}
\begin{split}
\cF_{x^0 x^0} = & \, \frac{1}{x^0} \left[ 2 N_{\La\Si} \,
A^\La \, A^\Si - \frac{(v^1)^2}{(x^0)^2} \, K(z, \zb) - 4 \, \Delta \right] \, , \\
{\mathcal F}_{x^0 v^\La} = & \, \frac{v^1}{2(x^0)^2}\bar
z^\Si N_{\La\Si} -\frac{1}{v^1}(\partial_\La N_{\Si\Xi}) \, A^\Si \, A^\Xi -\frac{2ic}{v^1}\left(1+\ln\, (x^0)\right) \left(\partial_\La\tF\right) \, , \\
\cF_{x^0 x^\La} = & - \, \, \frac{1}{x^0} \, N_{\La\Si} \, A^\Si \, , \\
{\mathcal F}_{x^\La v^\Si} = & \, \frac{1}{v^1}(\partial_\La N_{\Si\Xi}) \, A^\Xi \, , \\
\cF_{x^\La x^\Si} = & \, \frac{1}{2(x^0)} N_{\La\Si} \, .
\end{split}
\ee
Here $\partial_\La\equiv\partial /\partial v^\La$ and we have used
the homogeneity properties of $K(v, \vb)$, $N_{\La\Si}$, $D(v)$,
$\partial_\La N_{\Si\Xi}$, and $\partial_\La\tF(v)$ to write these
as functions of $z, \zb$ by extracting appropriate powers of
$v^1$. Furthermore, we defined
\be
  \dd \equiv \frac{ic}{2} \, [D(z)-\bar D(\bar z)] \, ,
\ee
and we will use $ K \equiv K(z, \zb) = z^{\Lambda} N_{\Lambda
\Sigma} \zb^{\Sigma}$ from now on.

In order to fix the remaining superconformal symmetry we still
need to implement the embedding equation \eqref{eq:D-eq}. Setting
$\kappa^2 = 1/2$ and substituting the derivatives \eqref{eq:qder},
eq.\ \eqref{eq:D-eq} is easily solved for $v^1$:
\be \label{5.8} v^1 = \sqrt{\frac{4x^0(1+\dd x^0)}{K}} \, . \ee
This relation expresses $v^1$ in terms of $x^0, x^\Lambda,
z^a, \zb^a$. Together with $A^\Lambda$, these provide the coordinates 
on the manifold
$\cM_{\rm TM}$. The $SU(2)$-gauge \eqref{eq:SU2-eq},
eq.\ \eqref{5.8} fixes the
  superconformal transformations $SU(2)$ and dilatations.
%
\subsection{Constructing the tensor multiplet lagrangian}
\label{SB.2}
%
Our next task is to compute the $SU(2)$ currents \eqref{vecS} and
\eqref{vecT} together with the inverse of the matrix $M$, eq.\
\eqref{1.2}. The former are obtained by substituting the
derivatives \eqref{eq:qder} into the general expressions for $\vec{S}_\mu$ and $\vec{T}_\mu$ and read
\be \label{eq:ST1} \vec S_\mu = \left[ \begin{array}{c} - i
\left(\frac{x^0(1+\dd x^0)}{K}\right)^{1/2} \, z^\La
N_{\La\Si} \partial_\mu A^\Si \\
i \left(\frac{x^0(1+\dd x^0)}{K}\right)^{1/2} \, \bar z^\La
N_{\La\Si}
\partial_\mu A^\Si \\ -2i \frac{(1+\dd x^0)}{K}\left(\partial_a K
\partial_\mu z^a - \partial_{\bar{a}} K \partial_\mu \bar z^{\bar{a}}\right) \end{array} \right]\,,
\ee
and
\be \label{eq:ST2} \vec T_\mu = \left[
\begin{array}{c} \left(\frac{1+\dd x^0}{x^0 K}
\right)^{1/2}\left(z^\La
N_{\La\Si} E_\mu^\Si - 2 z^\La N_{\La\Si} A^\Si E_\mu^0 \right) \\
\left(\frac{1+\dd x^0}{x^0 K} \right)^{1/2} \left(\bar z^\La
N_{\La\Si} E_\mu^\Si - 2 \bar z^\La N_{\La\Si} A^\Si E_\mu^0 \right) \\
- \frac{4}{x^0}(1+2\dd x^0) E_\mu^0 \end{array} \right] \, , \ee
respectively. Computing the matrix $M$, eq.\ \eqref{1.2}, we find
that it is block-diagonal with  entries $M_{\pm 3} = 0$. The
determinant of the $2\times2$-block is denoted by
\be\label{A.7}
  |\tilde M| = \frac{(1+\dd x^0)^2}{K^2}(zNz)(\bar z N \bar z)
-\dd^2(x^0)^2 \, , \ee
and it is easy to show that this coincides with \eqref{mtilde}.
The matrix $M^{-1}$ is found as
\be \label{eq:M} M^{-1} = \left[ \begin{array}{ccc} - \frac{(1+\dd
x^0)}{|\tilde M|K}(\bar z N \bar z) & \frac{\dd x^0}{|\tilde M|} &
0
\\ \frac{\dd x^0}{|\tilde M|} & - \frac{(1+\dd x^0)}{|\tilde
M|K}(zNz) & 0 \\ 0 & 0 & \frac{1}{4(1+\dd x^0)}
\end{array} \right] \, .
\ee
Here $zNz = z^\Lambda N_{\Lambda \Sigma} z^\Sigma$, etc.

The next step then consists in substituting these results into the
lagrangian \eqref{1.4} and employing the gauge fixing conditions
\eqref{eq:D-eq} and \eqref{5.8}. In order to make this calculation
more traceable, we take advantage of the fact that the gauge-fixed
lagrangian naturally splits into three sectors where the two
space-time derivatives act on scalar-scalar (SS), tensor-tensor
(TT) and scalar-tensor (ST) fields, respectively. We will now
consider these sectors in turn.
\subsubsection{The scalar-scalar sector}
The part of the lagrangian \eqref{1.4} where the two space-time
derivatives act on scalar fields is given by
\be \label{A.9} e^{-1} \cL^{\rm SS} = \cF_{x^I x^J} \left(
\partial_\mu v^I
\partial^\mu \vb^J + \tfrac{1}{4} \partial_\mu x^I \partial^\mu
x^J \right) - \tfrac{1}{2} \vec{S}_\mu M^{-1} \vec{S}^\mu \, . \ee
Substituting $M^{-1}$ from eq.\ \eqref{eq:M} and $\vec{S}_\mu$
given in \eqref{eq:ST1}, this becomes
\be\label{eq:p1a}
\begin{split}
e^{-1} \cL^{\rm SS}_{\rm qc} = & \,
- \frac{1 + 2 \dd x^0}{2 (x^0)^2 \left( 1 + \dd x^0 \right)} \,
(\partial_\mu x^0)^2 + 2 (1+\dd x^0) {\mathcal G}_{a \bar b}
\partial_\mu z^a \partial^\mu \bar z^{\bar
b } \\
& \, + 2 \, x^0 \, g_{\La\Si} \, \partial_\mu A^\La \partial^\mu
A^\Si
  - \frac{c^2(x^0)^2}{8(1+\dd x^0)}\left(\partial_a \tF
\partial_\mu z^a - \partial_{\bar a} \bar\tF \partial_\mu\bar z^{\bar a}\right)^2 \, .
\end{split}
\ee
Here $g_{\La \Si}$ is given by
\be
\begin{split}
g_{\La\Si} = & \, \frac{1}{4}\left[ N_{\La\Si}-\frac{(1+\dd
x^0)^2}{|\tilde M| K^2}
\left[(zNz)(N\bar z)_\La(N\bar z)_\Si + (\bar z N \bar z)(Nz)_\La(Nz)_\Si\right]\right] \\
& \,  - \left(\frac{\dd}{2}\right) \left(\frac{x^0 \, (1+\dd
x^0)}{|\tilde M|K} \right) (N z)_\La(N\bar z)_\Si \, ,
\end{split}
\ee
and the matrix $\cG_{a \bar{b}}$, defined in \eqref{eq:KM},
appears as a subsector of\footnote{This is completely analogous to
taking the superconformal quotient for vector multiplets
\cite{conformalcalculus} where the choice $z^{\Lambda} =
v^{\Lambda}/v^1$ corresponds to introducing special coordinates.}
\be {\mathcal G}_{\Lambda \bar \Sigma} =
  \partial_\Lambda
\partial_{\bar \Sigma} \, \ln \, K \, = \, \frac{1}{K}\left[ N_{\Lambda \bar \Sigma} -
\frac{1}{K}(Nz)_\Lambda (N\bar z)_{\bar \Sigma} \right] \, . \ee
We then construct the matrix
\be
\begin{split}
{\mathcal M}_{\La\Si} =  -i\bar F_{\La\Si} - \frac{(1+\dd
x^0)}{|\tilde M|K}\left[\frac{(1+\dd x^0)}{K}(\bar z N \bar
z)(Nz)_\La (Nz)_\Si + (\dd x^0)(Nz)_\La(N\bar z)_\Si \right] \ .
\end{split}
\ee
It can be rewritten as in \eqref{eq:def} and satisfies the
relation
\be \label{A.14} g_{\La\Si} = \tfrac{1}{4} \left( \cM + \bar{\cM}
\right)_{\La\Si} \, . \ee
This decomposition is motivated by the classical limit where, upon
setting $c = 0$, $\cM_{\La\Si}$ reduces to $\cN_{\La\Si}$ given in
\eqref{eq:cN}. This implies that $g_{\La\Si}|_{c = 0} =
\tfrac{1}{4} \left( \cN + \bar{\cN} \right)_{\La\Si}$ precisely
reproduces the scalar kinetic term of $(\partial_\mu A)^2$ in the
classical tensor multiplet lagrangian \eqref{classical}.

Using \eqref{A.14} and the definition $\tilde{A}_{\mu} = i \left[
\partial_a D(z) \partial_\mu z^a - \partial_{\bar a} \bar{D}(\zb)
\partial_{\mu} \bar{z}^{\bar a} \right]$ given in \eqref{4.11} the
lagrangian \eqref{eq:p1a} becomes
\be\label{eq:p1}
\begin{split}
e^{-1} \cL^{\rm SS}_{\rm qc} = & \,
- \frac{1 + 2 \dd x^0}{2 (x^0)^2 \left( 1 + \dd x^0 \right)} \,
(\partial_\mu x^0)^2 + 2 (1+\dd x^0) {\mathcal G}_{a \bar b}
\partial_\mu z^a \partial^\mu \bar z^{\bar
b } \\
& \, + \tfrac{1}{2} \, x^0 \, \left( \cM + \bar{\cM}
\right)_{\La\Si} \, \partial_\mu A^\La \partial^\mu A^\Si
  + \frac{c^2(x^0)^2}{8(1+\dd x^0)} \tilde{A}_{\mu} \, \tilde{A}^{\mu} \, .
\end{split}
\ee
\subsubsection{The tensor-tensor sector}
The part of the TM lagrangian \eqref{1.4} where the two space-time
derivatives act on tensor fields reads
\be
\begin{split}
e^{-1} \cL^{\rm TT} = & \, -{\mathcal F}_{x^I x^J} \, E_\mu^I \,
E^{J\mu} - \frac{1}{2} \, \vec T_\mu \, (M^{-1}) \, \vec T^\mu \,
.
\end{split}
\ee
Substituting the derivatives \eqref{eq:qder} together with
$M^{-1}$ and $\vec T_{\mu}$ given above it is straightforward to
rewrite this expression as
\be\label{eq:p2} e^{-1} \cL^{\rm TT}  = 2 \, \cT_{IJ}^{\rm qc} \,
E_\mu^I \, E^{J\mu} \, , \ee
with $\cT_{IJ}^{\rm qc}$ being the symmetric matrix
\be \label{A.18} \cT_{IJ}^{\rm qc} = \left[
\begin{array}{cc}
\frac{1 + 2 \Delta x^0}{(x^0)^2 \left(1 + \Delta x^0 \right)} + 4
\cT_{\La\Si}^{\rm qc} \, A^\La \, A^\Si   &
- 2 \, \cT_{\La\Si}^{\rm qc} \, A^\Si \\
- 2 \, A^\La \, \cT_{\La\Si}^{\rm qc}   & \cT_{\La\Si}^{\rm qc}
\end{array}
\right] \, , \ee
and
\be\label{eq:B.17}
\begin{split}
  \cT_{\La\Si}^{\rm qc} =
& \, - \frac{1}{4x^0}\left[N_{\La\Si}-\frac{(1+\dd x^0)^2}{|\tilde
M| K^2} \Big[(zNz)(N\bar z)_\La(N\bar z)_\Si + (\bar z N \bar
z)(Nz)_\La(Nz)_\Si \Big] \right]\\
& \, - \left(\frac{\dd}{4}\right)\frac{1+\dd x^0}{|\tilde
M|K}\left[(Nz)_\La(N \bar z)_\Si+(N\bar z)_\La (Nz)_\Si\right] \,
.
\end{split}
\ee
This expression can be rewritten as \eqref{eq:cTqc} in the main
text.

\subsubsection{The scalar-tensor sector}
Finally, we have the sector where the two space-time derivatives
act on a scalar and a tensor field. In this subsector eq.\
\eqref{1.4} gives rise to
\be e^{-1} \cL^{\rm ST} = iE_\mu^I( {\mathcal F}_{v^I
x^J}\partial_\mu v^I - {\mathcal F}_{\bar v^I x^J} \partial^\mu
\bar v^J) - \vec S_\mu (M^{-1})\vec T^\mu \, . \ee
Substituting the results \eqref{eq:M}, \eqref{eq:ST1} and
\eqref{eq:ST2} this becomes
\bea \label{eq:STint}
e^{-1} \cL^{\rm ST} & = & \, \, \partial_\mu \left(
F_{\La\Si}+\bar F_{\La\Si} \right) \, \left( A^\La A^\Si E^{0\mu}
- A^{\La} \, E^{\Si \mu} \right) \\ \nonumber && -  \frac{i(1+\dd
x^0)^2}{|\tilde M|K^2} \big[(zNz)(N\bar z)_\La(N\bar z)_\Si -
\mbox{c.c.} \big] \left(  2 A^{\La} \partial_\mu A^{\Si} E^{0\mu}
- E_\mu^\La\partial^\mu A^\Si \right) \\ \nonumber && + i \dd x^0
\, \frac{(1+\dd x^0)}{K} \left[(Nz)_\La(N\bar z)_\Si - (N\bar
z)_\La (N z)_\Si\right] \left(  2 A^{\La} \partial_\mu A^{\Si}
E^{0\mu}  - E_\mu^\La\partial^\mu A^\Si \right) \\ \nonumber && -
\frac{2i\dd}{K}\left[(N \bar z)_\La\partial^\mu z^\La - (Nz)_\La
\partial^\mu \bar z^\La\right]E_\mu^0 - \frac{2c}{x^0} \, \left(
\tF(z) + \bar\tF(\bar z) \right)  E_\mu^0 \, \partial^{\mu} x^0 \,
.
\eea
Here we have used the homogeneity property of $F_{\La\Si}$ to
rewrite space-time derivatives acting on $z^\La, \bar{z}^{\La}$ as
space-time derivatives of $F_{\La\Si}$ in the first line.
Furthermore, the last term was integrated by parts making use of
the Bianci identity for the tensor field strength, $\partial^\mu
E_\mu^I=0$.

Partially integrating the first term in \eqref{eq:STint}, we find
that the first three lines are all proportional to the combination
$\left(  2 A^{\La} \partial_\mu A^{\Si} E^{0\mu}  -
E_\mu^\La\partial^\mu A^\Si \right)$. The prefactor of this term
can then be reexpressed in terms of the matrix $\cM_{\La\Si}$
defined in \eqref{eq:def} by noting that
\be
\begin{split}
i \left( \cM - \bar{\cM} \right)_{\La\Si} = & \, F_{\La\Si}+\bar F_{\La\Si} \\
  & \, + i\frac{(1+\dd x^0)^2}{|\tilde M|K^2}\big[(zNz)(N\bar z)_\La(N\bar z)_\Si-(\bar z N \bar
z)(Nz)_\La(Nz)_\Si\ \big]\\
& \, - i \dd x^0 \frac{(1+\dd x^0)}{K}\left[(Nz)_\La(N\bar z)_\Si
- (N\bar z)_\La (N z)_\Si\right]  \, .
\end{split}
\ee
Recalling the definition of the K\"ahler connection
from eq.\ \eqref{eq:def} the expression \eqref{eq:STint} can then
be concisely written as
\be\label{eq:p3} e^{-1} \cL^{\rm ST}_{\rm qc} = i \left({\mathcal
M}-\bar{\cM}\right)_{\La\Si} \left[E_\mu^\La\partial^\mu A^\Si - 2
E_\mu^0 A^\La \partial^\mu A^\Si\right]
  \, - 2\dd A_\mu E^{0 \mu}
- \frac{2c}{x^0} \, \left( \tF + \bar\tF \right)  E_\mu^0
\partial^{\mu} x^0 \, . \ee

Summing the scalar-scalar \eqref{eq:p1}, tensor-tensor
\eqref{eq:p2}, and scalar-tensor contributions \eqref{eq:p3} then
yields the deformed tensor multiplet lagrangian
\bea \label{TML}
e^{-1} \cL^{\rm TM}_{\rm qc} & = & \,
- \frac{1 + 2 \dd x^0}{2 (x^0)^2 \left( 1 + \dd x^0 \right)} \,
(\partial_\mu x^0)^2 + 2 (1+\dd x^0) {\mathcal G}_{a \bar b}
\partial_\mu z^a \partial^\mu \bar z^{\bar
b } \\ \nonumber && \, + \tfrac{1}{2} \, x^0 \, \left( \cM +
\bar{\cM} \right)_{\La\Si} \, \partial_\mu A^\La \partial^\mu
A^\Si
  + \frac{c^2(x^0)^2}{8(1+\dd x^0)} \tilde{A}_{\mu} \, \tilde{A}^{\mu}
  + 2 \, \cT_{IJ}^{\rm qc} \, E_\mu^I \, E^{J\mu} \\ \nonumber
&& + i \left({\mathcal M}-\bar{\cM}\right)_{\La\Si}
\left[E_\mu^\La\partial^\mu A^\Si - 2 E_\mu^0 A^\La \partial^\mu
A^\Si\right]
   - 2\dd A_\mu E^{0 \mu}
- \frac{2c}{x^0}  \left( \tF + \bar\tF \right)  E_\mu^0
\partial^{\mu} x^0  .
\eea
Here $\cT_{IJ}^{\rm qc}$ is given in eq.\ \eqref{A.18}. Carrying
out the coordinate transformation $x^0 = {\rm e}^{-\phi}$ this
expression then gives rise to the deformed TM lagrangian
\eqref{onelooptensorlagrangian}. This result then completes the
derivation of the results presented in Section \ref{SS:5.2}.

We end this subsection by taking the classical limit of
\eqref{TML} by setting $c = \dd = 0$. In this case we have
\be {\mathcal M}_{\La\Si}|_{c = 0} = \cN_{\La\Si} \ee
and the matrix appearing in the tensor kinetic term simplifies to
\be \cT_{IJ}^{\rm cl} = \frac{1}{x^0} \, \left[
\begin{array}{cc}
\frac{1}{x^0} - \left( \cN + \bar{\cN} \right)_{\La\Si} A^\La
A^\Si
& \frac{1}{2} \, A^\La \, \left( \cN + \bar{\cN} \right)_{\La\Si} \\
\frac{1}{2}  \, \left( \cN + \bar{\cN} \right)_{\La\Si} \, A^\Si &
- \tfrac{1}{4}  \left( \cN + \bar{\cN} \right)_{\La\Si}
\end{array}
\right] \, . \ee
The resulting classical tensor multiplet lagrangian is then
precisely given by \eqref{classical}.

\subsection{Dualizing to hypermultiplets}
\label{SB.3}
After establishing the deformed TM lagrangian \eqref{classical} we
can now construct the quantum corrected HM lagrangian by dualizing
the tensor fields into scalars. For this purpose we add the
Lagrange multipliers introduced in eq.\ \eqref{eq:LMP}, 
\be e^{-1}{\mathcal L}^{\rm LM}=2(\partial_\mu w_0)E^{0\mu}-
(\partial_\mu B_\La)E^{\La\mu} \, , \ee
to the deformed TM lagrangian and compute the equations of motion
for the tensor fields. Using the notation
$E^I_\mu=\{E_\mu^0,E_\mu^\Lambda\}$, we find
\be \label{qceom} E_\mu^I = -\tfrac{1}{4}\left({\mathcal T}^{\rm
qc}\right)^{IJ}\,{\cJ}_{J\mu} \, . \ee
Here
\be ({\mathcal T}^{\rm qc})^{IJ} =
\frac{(x^0)^2}{f}\left[\begin{array}{cc} 1 & 2 A^\Si \\ 2 A^\La &
\frac{f}{(x^0)^2}  (\cT^{\rm qc})^{\La\Si} + 4 A^\La A^\Si
\end{array}\right] \, , \quad f = \frac{1 + 2 \dd x^0}{1 + \dd
x^0} \, , \ee
and $(\cT^{\rm qc})^{\La\Si}$ denotes the inverse of
\eqref{eq:B.17}. Furthermore, we have introduced the currents
coupling to the tensor field strength $E^{\La \mu}$ and $E^{0
\mu}$
\be \label{btildes}
\begin{split}  \cJ_{\La\mu} = & \, -  (\partial_\mu B_\La)+i({\mathcal M} -
\bar{\mathcal M})_{\La\Si}\partial_{\mu} A^\Si \, , \quad \mbox{and} \\
\cJ_{0\mu} = & \, 2 (\partial_\mu w_0) - 2 i \, ({\mathcal M} -
\bar{\mathcal M})_{\La\Si} A^\La\partial_\mu A^\Si - 2 \dd A_\mu -
\frac{2c}{x^0} (D(z)+\bar D(\bar z))(\partial_\mu x^0) \, ,
\end{split}
\ee
respectively. The dual deformed hypermultiplet lagrangian is then
obtained by using the equations of motion \eqref{qceom} to
eliminate the tensor fields from  ${\mathcal L}_{\rm qc}^{\rm
TM}+{\mathcal L}^{\rm LM}$. It reads
\be\label{eq:B30}
\begin{split}
e^{-1}{\mathcal L}^{\rm HM}_{\rm qc} = & \, - \frac{1 + 2 \dd x^0
}{2(x^0)^2 (1 + \dd x^0 )}(\partial_\mu x^0)^2
+ 2(1+\dd x^0) {\mathcal G}_{a \bar b }\partial_\mu z^a\partial^\mu \bar z^{\bar b} \\
& \, - \frac{1}{8}({\mathcal T}^{\rm qc})^{IJ} \cJ_{I\mu}
\cJ_J^\mu + \frac{x^0}{2}({\mathcal M}+\bar{\mathcal
M})_{\La\Si}\partial_\mu A^\La\partial^\mu A^\Si
    +\frac{c^2(x^0)^2}{8(1+\dd x^0)}\tilde A_\mu\tilde A^\mu \, .
\end{split} \ee
  Upon expanding the terms containing
$\cJ_I^\mu$ using the definitions \eqref{btildes} and setting $x^0
= {\rm e}^{-\phi}$ this expression gives rise to the deformed
hypermultiplet lagrangian \eqref{finallagrangian}. Note that in
the classical limit $c = 0$ \eqref{eq:B30} reduces to the
classical hypermultiplet lagrangian obtained from the c-map
\eqref{eq:cmap} with, as usual, $w_0$ given by \eqref{eq:ct}.

\subsection{The universal hypermultiplet example}
\label{SB.4}
In order to wrap up this section we now apply the formalism
outlined in the previous subsections to the universal
hypermultiplet. This HM geometry arises from compactifying type
IIA strings on a rigid CY$_3$ ($h^{1,2} = 0$) and consists of a
single (the universal) hypermultiplet. Classically this
hypermultiplet parameterizes the manifold
\be \cM_{\rm UHM} = \frac{SU(2,1)}{SU(2) \times U(1)} \, , \ee
and the string loop corrections to this space have recently been
found in \cite{AntoniadisUH}. It is thus interesting to compare
our results for generic CY$_3$ compactifications to the ones
obtained in the rigid limit.

In our conventions, the special K\"ahler geometry underlying the
classical universal hypermultiplet is determined by the
holomorphic prepotential
\be\label{A.31} F^{\rm UHM}(v) = - i (v^1)^2 \, . \ee
Furthermore, since rigid CY$_3$ do not have any complex structure
moduli, the function $D(v)$ appearing in the loop-correction term
has to be an imaginary constant. Without loss of generality we can
set $D(v) = -i$ since any rescaling of $D(v)$ can be absorbed into
a rescaling of the constant $c$.

Starting from the prepotential \eqref{A.31} it is straightforward
to calculate the objects specifying the corresponding special
K\"ahler geometry \eqref{2.3}
\be N_{11} = 4 \; , \quad K(v,\vb) = 4 v^1 \vb^1 \; , \quad K(z,
\zb) = 4 \, . \ee
Here we used the definition of the inhomogeneous coordinates
\eqref{inhomogeneous} to obtain the last expression. Substituting
these quantities into the general formula for $\cF(x,v,\vb)$, eq.\
\eqref{result}, we obtain
\be \cF^{\rm UHM}( v, \vb ,x ) = \frac{1}{x^0} \, \left( (x^1)^2 -
2 \, v^1 \, \vb^1 \, \right) - 4 \, c \, x^0 \, \ln(x^0) \, . \ee
Note that this result agrees with the one obtained from evaluating
the contour integral for $H^{\rm UHM}(\eta)$, eq.\ \eqref{qcuhm},
in the partial gauge $v^0 = 0$ and for a contour encircling the
origin.

In order to obtain the loop-corrected tensor multiplet lagrangian
corresponding to the universal hypermultiplet we can then proceed
by evaluating the definitions \eqref{eq:cTqc} and \eqref{eq:def}
for the specific prepotential \eqref{A.31}. Using that \eqref{A.7}
becomes
\be |\tilde M| = 1 + 2 c x^0 \, , \ee
these read
\be \label{A.36}
\begin{split}
\cT^{\rm qc}_{11} = \frac{1}{x^0 \, (1 + 2 c x^0 )} \, , \quad
\cM_{11} = -2 \left( 1 + 2 c x^0 \right) \, .
\end{split}
\ee
Furthermore, the connections \eqref{4.11} and \eqref{4.12} vanish
in the absence of complex structure moduli.

Substituting these results into the general lagrangian \eqref{1.4}
we obtain a double tensor multiplet description of the quantum
corrected universal hypermultiplet\footnote{Since the universal
hypermultiplet possesses different and inequivalent pairs of
commuting isometries which can be used to dualize scalars into
tensor fields, the description of the universal hypermultiplet in
terms of a double tensor multiplet is not unique \cite{HKC1}, so
strictly speaking a ``universal double tensor multiplet'' does not
exist.}
\be e^{-1} \cL^{\rm DTM}_{\rm qc} = - \tfrac{1}{2 (x^0)^2} \,
\tfrac{1 + 2 c x^0}{1 + c x^0} (\partial_{\mu} x^0)^2  + 2 x^0
\left( 1 + 2 c x^0 \right) ( \partial_\mu A)^2 + 2 \,
\cT_{IJ}^{\rm DTM} \, E^I_\mu \, E^{J \mu} \, , \ee
where $\cT_{IJ}^{\rm DTM}$ is given by
\be
\begin{split}
\cT_{IJ}^{\rm DTM} = & \, \frac{1}{x^0 (1 + 2 c x^0)} \left[
\begin{array}{cc}
  \left(  \frac{1}{x^0} \, \frac{ (1 + 2cx^0)^2}{1 + c x^0} + 4 \, A^2 \right) & - 2 \, A \\
- \, 2 A & 1 \\
\end{array}
\right] \, . \\
\end{split}
\ee
and we have dropped the (redundant) index from $A^1$.

We note that taking the classical limit $c = 0$ and using the
coordinate transformation
\be\label{eq:trafo} x^0 = {\rm e}^{- \phi} \, , \quad A =
\tfrac{1}{2} \, \chi \, , \quad E^{I\mu} = \frac{1}{2} H^{\mu I}
\, , \ee
this is precisely the double tensor multiplet lagrangian obtained
in ref.\ \cite{UHMsuper}
  \be e^{-1} \cL^{\rm DTM} = - \, \frac{1}{2}
\partial^{\mu} \phi
\partial_{\mu} \phi - \frac{1}{2} {\rm e}^{-\phi} \partial^{\mu}
\chi \partial_{\mu} \chi + \frac{1}{2} M_{IJ} H^{\mu I} H_{\mu}^J
\ee
where
\be M_{IJ} = {\rm e}^{\phi} \left[
\begin{array}{cc}
{\rm e}^{\phi} + \chi^2 & - \chi \\
- \chi & 1
\end{array}
\right] \, . \ee

In order to obtain the quantum corrected HM lagrangian, we
substitute the equations \eqref{A.36} into
\eqref{finallagrangian}. Denoting $f = \frac{1 + 2 c x^0}{1 + c
x^0}$ as in eq.\ \eqref{eq:def} the result becomes
\be\label{A.42}
\begin{split}
e^{-1} \cL^{\rm UHM}_{\rm qc} = & \, - \frac{f}{2 (x^0)^2} \,
(\partial_\mu x^0 )^2
- \tfrac{1}{2} x^0 \, (1 + 2 c x^0) \left( 4 (\partial_\mu A)^2 + \tfrac{1}{4} (\partial_\mu B)^2 \right) \\
&  \, - \frac{(x^0)^2}{2 f}  \left( \partial_\mu w_0 - A
\partial_\mu B \right)^2  \, .
\end{split}
\ee
One can easily check (using Mathematica) that the metric of this
non-linear sigma model is Einstein with Ricci curvature $R = -6$.

We now compare the lagrangian \eqref{A.42} to the result for the
loop corrected universal hypermultiplet moduli space found in
\cite{AntoniadisUH}. There the corresponding line element was
given in Calderbank-Pedersen form \cite{CalderbankPedersen} as:
\be \label{B.15}
\begin{split}
\frac{1}{2}\, \rmd s^2 = & \, \frac{(\rho^2+\hat
\chi)}{(\rho^2-\hat\chi)^2}\left[(\rmd
\rho)^2+(\rmd\eta)^2+\frac{(\rmd\phi)^2}{4}\right]+\frac{\rho^2}{(\rho^2-\hat\chi)^2(\rho^2+\hat\chi)}\left[\rmd\psi+\eta\,\rmd\phi\right]^2
\, .
\end{split}
\ee
Moreover, the value of the constant $\hat \chi$ 
was determined to be
\be\label{B.17} \hat\chi \equiv  - \, \frac{4 \, \zeta(2) \,
\chi(X)}{(2 \pi)^3} = - \frac{1}{6 \pi} \left( h^{1,1} - h^{1,2}
\right) \, . \ee

In order to match \eqref{A.42} with \eqref{B.15} we perform the
coordinate transformation
\be x^0 = \left( \rho^2 -c \right)^{-1} \; , \quad A = \eta \; ,
\quad B = 2 \phi \; , \quad w_0 = - 2 \psi \, , \ee
which brings \eqref{A.42} into Calderbank-Pedersen form~\footnote{We use conventions in which the lagrangian takes the form $e^{-1}{\cal L}=-R-{\rm d}s^2$.}:
\be \label{B.15us}
\begin{split}
\frac{1}{2}\, \rmd s^2 = & \,
\frac{(\rho^2+c)}{(\rho^2-c)^2}\left[(\rmd
\rho)^2+(\rmd\eta)^2+\frac{(\rmd\phi)^2}{4}\right]+\frac{\rho^2}{(\rho^2-c)^2(\rho^2+c)}\left[\rmd\psi+\eta\,\rmd\phi\right]^2
\, .
\end{split}
\ee
Comparing the expressions \eqref{B.15} and \eqref{B.15us} allows
us to read off the value of the constant $c$ as
\be \label{cconstant} c=\hat\chi \equiv  - \, \frac{4 \, \zeta(2)
\, \chi(X)}{(2 \pi)^3} =
  - \frac{1}{6 \pi} \left( h^{1,1} - h^{1,2} \right) .
\ee
This is also the value we have used in Section \ref{fcc}.


%
%
\end{document}